\documentclass[11pt]{article}\usepackage{amssymb,amsmath,amsfonts,eurosym,geometry,ulem,graphicx,caption,color,setspace,sectsty,comment,footmisc,caption,natbib,pdflscape, array, graphicx, subcaption, float, booktabs, appendix, tabularx, mathtools, threeparttable, lscape, enumerate, bbm, tikz-qtree, rotating, color, chngcntr}
\usepackage{hyperref}
\hypersetup{
colorlinks=true,
citecolor=blue,
linkcolor=blue,
filecolor=blue,      
urlcolor=blue,
}

\DeclareMathOperator*{\argmin}{arg\,min}
\newcommand{\bX}{\mathbf{X}}

\newcolumntype{L}[1]{>{\raggedright\let\newline\\arraybackslash\hspace{0pt}}m{#1}}
\newcolumntype{C}[1]{>{\centering\let\newline\\arraybackslash\hspace{0pt}}m{#1}}
\newcolumntype{R}[1]{>{\raggedleft\let\newline\\arraybackslash\hspace{0pt}}m{#1}}

\newcommand{\review}[1]{{\color{black} #1}}

\begin{document}

\begin{titlepage}
\title{Machine Learning for Zombie Hunting:\\ Predicting Distress from Firms' Accounts and Missing Values \footnotetext{\scriptsize{We are grateful to participants to the Bank of Italy/CEPR/EIEF conference on `Firm Dynamics and Economic Growth', to the Bank of England/King's College conference on `Modelling with Big Data and Machine Learning', to the Annual Conference of the JRC Community of Practice in Financial Research organized by the European Commission, and to the workshops on `Data Science for Impact Evaluation' jointly organized by KU Leuven and IMT School for Advanced Studies. We want to thank Tommaso Aquilante, Nicola Benatti, Kristina Bluwstein, Elena Cefis, Giulio Bottazzi, Dimitrios Exadaktylos, Nicol\'o Fraccaroli, Mahdi Ghodsi, Andreas Joseph, Francesca Lotti, Francesco Manaresi, Juri Marcucci, Andrea Mina, Chiara Osbat, Gianmarco Ottaviano, Giacomo Rodano, Andrea Roventini, Gabriele Rovigatti, Abhishek Samantray, Federico Tamagni, Francisco Queiro, Gias Uddin, Nicolas Woloszko and Nicol\'o Vallarano for their valuable comments.}}} 
\author{Falco J. Bargagli-Stoffi\thanks{\scriptsize{Mail to: \href{mailto://fbargaglistoffi@hsph.harvard.edu}{\color{blue}fbargaglistoffi@hsph.harvard.edu}. Harvard University, 677 Huntington Ave, Boston, MA 02115, United States.} } \and 
Fabio Incerti\thanks{\scriptsize Mail to: \href{mailto://fabio.incerti@imtlucca.it}{\color{blue}fabio.incerti@imtlucca.it}. Laboratory for the Analysis of Complex Economic Systems, IMT School for Advanced Studies, piazza San Francesco 19 - 55100 Lucca, Italy.} \and 
Massimo Riccaboni\thanks{\scriptsize Corresponding author, mail to: \href{mailto://massimo.riccaboni@imtlucca.it}{\color{blue}massimo.riccaboni@imtlucca.it}. Laboratory for the Analysis of Complex Economic Systems, IMT School for Advanced Studies, piazza San Francesco 19 - 55100 Lucca, Italy.} \and Armando Rungi\thanks{\scriptsize Mail to: \href{mailto://armando.rungi@imtlucca.it}{\color{blue}armando.rungi@imtlucca.it}. Laboratory for the Analysis of Complex Economic Systems, IMT School for Advanced Studies, piazza San Francesco 19 - 55100 Lucca, Italy.}}
\date{ 
} 

\maketitle
\vspace{-1cm}
\begin{abstract}
\footnotesize
\noindent In this contribution, we propose machine learning techniques to predict\textit{ zombie firms}. First, we derive the risk of failure by training and testing our algorithms on disclosed financial information and non-random missing values of 304,906 firms active in Italy from 2008 to 2017. Then, we spot the highest financial distress conditional on predictions that lies \review{above a threshold for which a combination of false positive rate (false prediction of firm failure) and false negative rate (false prediction of active firms) is minimized}. Therefore, we identify \textit{zombies} as firms that persist in a state of \review{financial distress, i.e., their forecasts fall into the risk category above the threshold for at least three consecutive years}. For our purpose, we implement \review{a gradient boosting algorithm (XGBoost) that exploits information about missing values. The inclusion of missing values in our predictive model} is crucial because patterns of undisclosed accounts are correlated with firm failure. Finally, we show that \review{our preferred machine learning algorithm} outperforms (i) proxy models such as Z-scores and the Distance-to-Default, (ii) traditional econometric methods, and (iii) other widely used machine learning techniques. We provide evidence that \textit{zombies} are on average less productive and smaller, and that they tend to increase in times of crisis. Finally, we argue that our application can help financial institutions and public authorities design evidence-based policies--- e.g., optimal bankruptcy laws and information disclosure policies.
\vspace{10pt}
\bigskip
\end{abstract}
\setcounter{page}{0}
\thispagestyle{empty}
\end{titlepage}
\pagebreak \newpage

\onehalfspacing

\section{Introduction} \label{sec:introduction}

In this paper, we propose machine learning techniques as suitable tools to make predictions about business failure when information about firm viability is partially undisclosed. Therefore, we define \textit{zombies} as firms that persist in a high-risk status because their predicted probability of failure \review{ is above a threshold for which a combination of the false positive rate (i.e., predicting a firm that did not fail as failed) and the false negative rate (i.e., predicting a firm that failed as not failed) is minimized}. We prove that the latter is also the distributional segment in which the probability of transitioning to a lower risk of failure is minimal. According to our machine learning approach, the Italian proportion of \textit{zombie} firms in the analysis period is between \review{1.5\% and 2.5\%}. \textit{Zombie} firms are on average \review{19\%} less productive and \review{17\%} smaller than viable firms. Interestingly, we find that \textit{zombies} are countercyclical, as their share increases in times of crisis and decreases in times of economic recovery.

The problem of identifying nonviable firms is important to scholars and practitioners, whether to assess the credit risk of an individual firm or to identify the part of an entire economy that is in trouble. \review{Originally, the notion of \textit{zombie firms} was associated with the phenomenon of ``\textit{zombie lending}", in which banks extend credit to otherwise insolvent borrowers. In some cases, \textit{zombie lending} is a deliberate strategy to avoid a bank's budget restructuring while only apparently complying with capital standards set by financial regulators \citep{bonfim2020site}, e.g., the case of Japanese banks in the 1990s \citep{peek2005unnatural,caballero2008zombie}. More recently, \citet{schivardi2021credit} studied Italian companies during the 2008 financial crisis and found that the misallocation of credit under \textit{zombie lending} increased the default rate of otherwise healthy firms while decreasing the default rate of nonviable firms. This was because undercapitalized banks could restrict lending to more viable projects to avoid disclosing nonperforming loans in their portfolios. Paradoxically, severely distressed companies can appear resilient in times of financial crisis thanks to continued access to financial resources.

From a more general perspective, recent studies introduce \textit{zombies} as firms that have persistent problems meeting their interest payments \citep{andrews2017confronting, mcgowan2018walking, banerjee2018rise, andrewspetroulakis}, thus expanding the original category to include firms that are in some form of financial distress. Based on proxy indicators from financial accounting, they show that \textit{zombies} account for a non-negligible share of modern economies--- up to 10\% of incumbent firms--- while absorbing up to 15\%, 19\%, and 28\% of the capital stock in countries such as Spain, Italy, and Greece, respectively. When market exit or restructuring delays occur, they drag down aggregate productivity by hindering the reallocation of resources in favor of healthier firms and preventing the entry of potentially more innovative and younger firms. It is therefore argued that identifying non-viable, financially distressed firms can be particularly useful in avoiding the misallocation of productive and financial resources.

Against this background, we argue that the empirical problem of assessing whether a firm is a \textit{zombie} is closely related to the more general problem of determining its credit risk. Ultimately, a \textit{zombie} is a non-viable firm that may escape bankruptcy despite its extreme financial distress--- i.e., despite scoring the highest credit risk. From another perspective, healthier firms are the furthest from bankruptcy and \textit{zombie} status. Traditionally, credit risk has been studied from the perspective of a financial firm, which must assess the health of a company using information, albeit limited, from financial books and public records.\footnote{The seminal reference is to a departure from the \citet{modiglianimiller} theorem, according to which capital structure should not be relevant to a company's value if there are no market frictions, including bankruptcy costs. Thus, a firm's ability to raise external financing should depend solely on the profitability of its investment projects. However, since financial market frictions cannot be eliminated, a firm's capital structure actually provides information about the profitability of the firm and its assets. See also the discussion of \citet{rajanzingales} for an international perspective} Thus, for decades, academics and practitioners have attempted to determine a firm's profitability after benchmarking exercises on firm-level indicators of financial constraints--- e.g., in estimating Z-scores \citep{altman1968financial, altman2000predicting}, Distance-to-Default \citep{merton1974pricing}, or investment- to- cash- flow sensitivity \citep{fazzarihubbardpetersen}. However, information on corporate viability may be incomplete due to strategic disclosure of relevant information and simplified financial records for certain categories of unlisted companies.


Thus, we propose a machine learning approach to predict credit risk and \textit{zombie} status from incomplete financial accounts. To show the potential of our approach, we work with a sample of 304,906 Italian firms over the period 2008-2017. Italy is a compelling example of a country that hosts a relevant share of inefficient firms that hinder the growth potential of the economy \citep{calligarisetal}.


The underlying intuition is simple: based on the experience of firms that failed in previous periods, we derive predictions about the risk of failure of active firms. Each time we compare the observed outcomes with the predicted outcomes, the algorithm updates and reduces the prediction errors in the next periods after processing new ``in-sample" information about the financial accounts. In the end, we obtain a probabilistic measure of the likelihood that a business will fail. Our machine learning framework improves upon existing benchmark models by leveraging a rich set of firm-level economic and financial indicators that potentially contain diverse information about both the firm's core economic activity and its ability to meet financial obligations.

Importantly, we find that emerging patterns of missing financial accounts are correlated with firm failure, possibly due to the fact that managers are more likely to conceal accounts when they are in financial distress. We provide evidence that most missing variables are often those that have been used as proxies for \textit{zombies} or financial constraints in the previous literature. Therefore, we implement \review{our missing-aware methodologies} incorporating patterns of undisclosed accounts \review{and test whether a substantial improvement in prediction occurs when missing data are properly handled}.
Using the missing values information as another predictor of outcome, we show that the best predictive algorithm in our setting, eXtreme Gradient Boosting (XGBoost) \citep{chen2016xgboost}, explains up to $0.97$ of the Area-Under-the-Curve (AUC), and its Precision-Recall (PR) performance reaches $0.76$. XGBoost outperforms credit scoring models (i.e., Z-score and Distance-to-Default models), standard econometric methods (i.e., logistic regression), and also other machine learning techniques (i.e., Classification and Regression Tree and Random Forests). 

We argue that asymmetric and undisclosed (missing) information is ubiquitous in corporate financial accounts. Therefore, simply omitting records with missing values would severely impair the search for predictors of \textit{zombies} and lead to a loss of precision and bias \citep{little2019statistical}. If the missing information is correlated with firm failures, excluding these firms would significantly reduce the number of failures in the sample and thus hinder the algorithm's learning for these instances. This problem could potentially be avoided by a missing data imputation approach that would allow the inclusion of these firms in the training sample. On the other hand, the missing information \textit{per se} constitutes relevant information to learn from failures. This may be the case if nonviable firms have the option of not disclosing information in their financial accounts. As a result, the mechanism in the data is Missing-Not-At-Random (MNAR), and the complete records are not a random sample of the population of interest. In this second case, if some companies consistently avoid disclosing part of their information, imputation of missing data would not be sufficient to solve the problem. Indeed, we know from previous literature that techniques for imputing data, such as mean or median imputation, can fail in the presence of MNAR because they distort the empirical distributions.\footnote{Further limitations of traditional approaches to imputing missing data are discussed in \citep{he2010multiple, white2018imputation, little2019statistical}.}}

\review{Based on the previous considerations, we choose an approach that can simultaneously incorporate missing data into our analysis while being robust to MNAR. We implement two methods specifically designed to deal with non-random patterns of missing data: XGBoost and Bayesian Additive Regression Trees with Missing Incorporated in Attributes (BART-MIA) \citep{kapelner2015prediction}. Although these methods differ in implementation, they have very similar routines for dealing with MNAR patterns.
XGBoost uses default directions, also known as block propagation \citep{josse2019consistency}, to group all incomplete observations and send them to one side of the tree. The MIA method \citep{twala2008good}, extended by \cite{kapelner2015prediction, JSSv070i04}, strengthens block propagation so that missingness can be used as an explicit feature to compute the best splits. According to \cite{josse2019consistency}, MIA can handle both informative and non-informative missing values.
Our results suggest that both XGBoost and BART-MIA effectively capture the direct influence of missing values as either implicit (block propagation) or explicit (MIA) predictors of the response variable. Finally, in the following analyzes, we show that XGboost has a significantly lower computational cost and relatively higher predictive power than BART-MIA. }

To shed more light on the information that contributes most to prediction, we use an interpretable framework that has its roots in game-theoretic Shapley values, introduced by \cite{strumbelj2010efficient}. Shapley values have recently been proposed to identify the economically meaningful nonlinearities learned by machine learning models \citep{buckmann2021interpretable}. We find that no single financial indicator predicts failure better than the ensemble of predictors used in our machine learning approach. 

Economically informative groups of variables have heterogeneous predictive power. In particular, indicators of firms' financial constraint or previously used indicators of \textit{zombies} are important. Nevertheless, they are of secondary importance when compared to information on firms' \review{financial accounts, indicators of corporate governance}, and the presence of non-random missing values. Our results confirm that machine learning techniques perform better than single indicators when incorporating as much valuable in-sample information as possible while updating each time there is new out-of-sample information because prediction errors dynamically decrease after independent tests that minimize the discrepancies between realized and predicted outcomes \citep{athey2018impact}.

Our framework is of particular interest to policymakers in designing optimal bankruptcy laws.\footnote{See also the suggestions by the European Directive 2012/30/EU, and the recent Italian Law on business failures on October 19, 2017, n. 155, which provides the legal basis for early notification of corporate crises to improve targeted interventions.} Tracking a company's bankruptcy risk allows all stakeholders, not just creditors, to understand whether there is an opportunity for restructuring and, if not, to prevent incumbent, albeit non-viable, firms from wasting additional economic resources. Evidence-based, interpretable methods are even more important after the recent pandemic crisis, as we believe that financial support must be targeted to companies that have a real chance of recovering and staying on their feet in normal times to avoid misallocation of resources.\footnote{For a first analysis of the impact of the pandemic crisis on Italian companies, see \citet{SchivardiRomano}}

\section{Data and preliminary evidence} \label{sec:data}

We obtain the financial accounts from the ORBIS database\footnote{ORBIS firm-level data \citep{orbis} have become a common source of global financial accounts. For previous use of this database, see \cite{gopinath2017capital} and \cite{cravinolevchenko}, among others. Coverage of smaller firms and some financial accounts may change across countries as national business registries impose different filing requirements, as observed in the validation exercises of \cite{kalemli2015construct} and \cite{gal2013measuring}. In the case of Italy, the original information provider for Italian financial accounts is CERVED, a credit rating agency. Bureau Van Dijk standardizes and translates the original financial accounts to make them comparable between countries. Note that, unlike other platforms of the same Bureau Van Dijk (e.g., AIDA or AMADEUS), ORBIS does not drop exiting firms in our analysis period. It supplements the financial accounts with other information from various sources on ownership, management and intellectual property rights, which we also use for predictions.}, compiled by the Bureau Van Dijk, for manufacturing firms active in Italy for at least one year from 2008-2017. Italy is a compelling case to study business failure and \textit{zombie firms}: it is a country where relatively inefficient firms hinder the economy's growth potential \citep{calligaris2016italy, bugamelli}, perpetuating geographic divergence \citep{rungi2019heterogeneous}, and are studied extensively by international organizations \citep{mcgowan2018walking, andrewspetroulakis}.

For our purpose, we use two main variables that help us identify business failure: the status of a firm and the date on which it becomes inactive. Table \ref{tab:italy_status} shows our sample coverage by firm status over in the period of analysis. We assume that a firm has failed in the first year if it is reported as ``bankrupt'', ``dissolved'', or ``in liquidation'', as in the original data. Overall, the share of exiting firms accounts for about 5.7\% of the total sample, which is close to the average official 6.3\% obtained by ISTAT, the national statistics office, for the same period.

\begin{figure}[H]
\centering
\caption{\review{Geographic coverage}}
\captionsetup{font=footnotesize}
\label{fig:map_logs}
\includegraphics[width=0.7\textwidth]{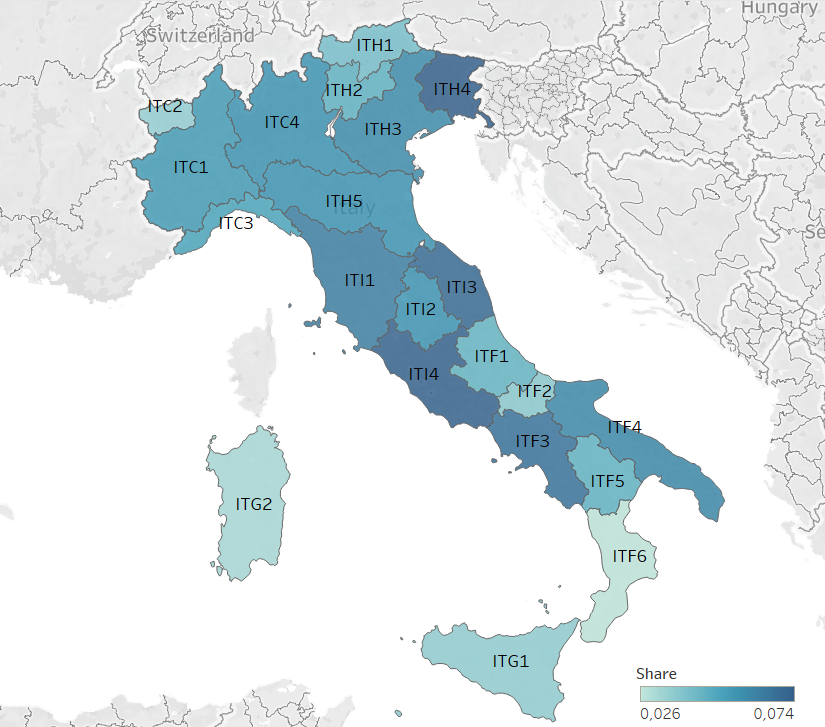}
\singlespacing
\caption*{\review{Proportion of business failures to the total number of firms for NUTS 2-digit regions. We assume that a firm fails when it is reported as bankrupt, dissolved, or in liquidation.}}
\end{figure}

\review{Figure \ref{fig:map_logs} shows the proportions of firm failures by NUTS 2-digit region. As expected, we find a higher concentration of failures in the north and center of the country, where economic density is higher. It is noteworthy that we fully represent the whole Italian territory since we detect business failures in every region during our period of analysis.}

\begin{table}[H]
\centering
\caption{Firms by status}
\label{tab:italy_status}
\resizebox{0.7\textwidth}{!}{%
\begin{tabular}{cccccc}
\toprule
Status     & Active & Bankrupted & Dissolved & In Liquidation & Total    \\
\midrule
Sample     & 287,586  & 1,533       & 8,540      & 7,221           & 304,906   \\
Percentage & 94.33\% & 0.50\%     & 2.80\%    & 2.37\%         & 100\% \\
\bottomrule \\
\end{tabular}
}
\end{table}

We use a set of economic and financial indicators to train our predictive models. The battery of predictors includes (i) original financial accounts at the firm level; (ii) widely used indicators to proxy firm-level financial constraints; (iii) indicators previously used to detect \textit{zombie firms}; (iv) indicators included as warnings of corporate crises in the recent Italian bankruptcy law.\footnote{A recent reform of the bankruptcy law (L. 155/2017 and DL. 14/2019) proposes an early warning system based on indicators identified by practitioners, the purpose of which is to identify companies in distress in time to intervene to preserve entrepreneurial capabilities and find a way out of the crisis. It delegated practitioners (in particular the National Association of Chartered Accountants: Consiglio Nazionale dei Dottori Commercialisti e degli Esperti Contabili ) to draw up a list of indicators that could help assess the state of crisis of a company} Each predictor we consider is described in detail in Table \ref{tab:table2} in the Appendix \ref{sec:appendix_a}.

Note that many indicators we select as predictors have been used in various frameworks to assess the extent to which a firm is in trouble. From our machine learning perspective, they cannot be interpreted as drivers of failure. It is sufficient that they contribute, albeit in a small way, to the assessment of a company's health. In our predictive framework, they might even border on multicollinearity, e.g., in the case of different measures of efficiency, liquidity, and solvency ratios. Since we are not interested in identifying a causal contribution to firm failure, high collinearity does not pose a problem for our predictions \citep{makridakis2008forecasting, shmueli2010explain}. On the contrary, we will discuss in Section \ref{sec:indicators} how, by construction, one cannot separate the empirical contribution of an indicator from the entire set in the context of a pure prediction problem. For this reason, the predictors we use should be considered as an inseparable ensemble, and a discussion of the statistical significance of the individual predictors is not relevant in our framework.

In addition to mandatory and basic information (volume of activity, profits, location, industry affiliation, ownership, and intellectual property rights), many other financial accounts have different patterns of missing values over time. \review{In the Appendix \ref{sec:appendix_a}, Figure \ref{fig:non-failing}, we visualize a map of missing values in our sample. The frequency of missing values affects all firms in the data, both active and failed, and it is more concentrated in the financial accounts of recent years. In addition, liquidations exhibit less pronounced patterns over time than the other failure categories}. After running a series of chi-squared tests (see Table \ref{tab:chi-squared}), we find a positive statistical relationship between the patterns we observe in the sample and the event of a firm's failure. In short, a firm is more likely to fail if a pattern of missing financial accounts is observed. The exercise performed in Table \ref{tab:logit_missing} clearly shows such correlations. We run a simple logistic regression using the observed failure of a firm as the dependent variable. We then add a binary regressor equal to one if the predictor is missing at least once in the three years prior to failure. Fixed effects by NUTS 3-digit region and NACE 2-digit industry are included. We report results for each predictor per row in Table \ref{tab:logit_missing}.

\begin{table}
\centering
\caption{Missing predictors and firms' failures} 
\label{tab:logit_missing}
\captionsetup{font=footnotesize}
    \resizebox{0.7\textwidth}{!}{%
        \begin{tabular}{lcccc}
            \toprule
            Missing predictor & Odds ratio & Std. Error & N. obs. & Pseudo R2\\
            \midrule
            Interest Coverage Ratio & 1.91*** & (0.30) & 304,906 & 0.026\\
            Interest Benchmarking & 1.29*** & (0.20) & 304,906 & 0.019\\
            value added & 2.37*** & (0.37) & 304,906 & 0.033\\
            Z-score & 2.55*** & (0.38) & 304,906 & 0.037\\
            Total Factor Productivity & 2.37*** & (0.39) & 304,906 & 0.034\\
            Profitability & 1.91*** & (0.30) & 304,906 & 0.027\\
            \bottomrule
        \end{tabular}}\\
\singlespacing
 \caption*{Odds ratios according to a logit specification in which the dependent variable is a firm failure and the binary regressor equals one if at least one missing value was found in the last three years. Fixed effects at the region and industry level. Errors are clustered by industry.}
    \end{table}

The above correlations are particularly relevant to the scope of our analyses. The main problem is sample selection when observations are selectively missing for some categories of firms. In this case, there are two potential sources of sample selection bias: (i) distressed firms \textit{vis \`{a} vis} firms that are not in distress, as the former may have the incentive to disclose less information than the latter; (ii) smaller companies firms \textit{vis \`{a} vis} bigger firms because the first are often exempted from a complete financial report, in accordance with Italian Regulation.\footnote{Under Italian civil law, companies that do not list financial activities on the stock exchange have the option to provide more aggregate financial reports if their size does not simultaneously exceed two of the following thresholds in one or two consecutive periods: i) 4,400,000 euros in total assets; ii) 8,800,000 euros in operating revenues; iii) 50 employees. A simplified financial statement always includes the most significant items in the first or second position digit of the aggregation.} Of course, the two sets of firms may overlap, as smaller companies may also be the ones that are proportionally more affected by financial difficulties. Using a missing-aware procedure, as described in Section \ref{sec:strategy}, allows us to consider both sources of sample selection when patterns of missing financial accounts emerge since the algorithm considers such patterns as another predictor of firm failure.

Interestingly, we note that most of the missing variables are also those used in previous work as proxies for \textit{zombies} or financial constraints. Take, for example, the case of the Interest Coverage Ratio (ICR), which is derived as the ratio between a firm's earnings before interest and taxes (EBIT) and its interest expense. If the ICR is less than one, \citet{bankofengland2013} assumes that a company is a \textit{zombie} because it is having trouble meeting its financial obligations. In our sample, we find that about 19\% of firms have an ICR smaller than one, but at the same time, there are 62.50\% firms whose ICR information is not available at all. Moreover, according to \citet{bankofkorea2013}, negative value added is the most appropriate indicator for evaluating a \textit{zombie} status, as it indicates that intermediate inputs have a higher market value than the firm's output. In the case of Italy, about 64.27\% of enterprises do not report their value added in at least one period, while about 3\% of them report a negative value. A negative value added is, of course, a more severe condition than a negative profit, since a company can make no profits without destroying economic value. Indeed, firms' profitability is at the heart of two similar proxies for \textit{zombies} used by \citet{schivardi2021credit} when comparing firm-level profits to an external benchmark. Repeating the same exercises, we find that about 3\% of Italian firms are distressed, while a large part of the sample (62.50\%) does not report any information on the predictor. Finally, both \citet{caballero2008zombie} and \citet{mcgowan2018walking} perform another benchmarking exercise, comparing the interest a firm pays to raise external finance with the cost opportunity to invest in alternative, safer investments. In our case, when we try to reproduce the same exercise with the yields of Italian government bonds with a ten-year maturity, we find that there is a high proportion (60.29\%) of companies for which we have no information in at least one period in which they were active.

We also include an estimate of Total Factor Productivity (TFP) as a predictor, following the methodology proposed by \cite{ackerberg2015identification}, to account for the simultaneity bias arising from \textit{ex-post} adjustments in combining factors of production. In this respect, firm-level TFP allows us to make predictions based on the ability to transform inputs and sell output in the market. Indeed, the relationship between financial constraints and productivity is one of the most debated issues \cite[see, among others,][]{aghionetal, ferrandoetal}. The simple assumption for our forecasting models is that less productive firms are the ones that have more difficulty surviving in the market. At the same time, \textit{zombie firms} have also often been defined in terms of (lack of) productivity \citep{mcgowan2018walking, andrews2017confronting, andrewspetroulakis, schivardi2020identifying}.\footnote{Please note how \citet{white2018imputation} recently highlighted the limitations of using missing imputation techniques for the variables used to calculate TFP in the US.} 

\section{Empirical strategy} \label{sec:strategy}

It is difficult to identify non-viable companies for obvious reasons. If financial accounts are in bad shape, one could argue that it is only a matter of time before they become more competitive if conditions are right. If the balance sheets are good, one could argue that the worst is yet to come because bad management decisions will show up later. Trivially, only the already bankrupt companies were certainly not viable at some point. Still, an outside observer will never know when that happened because the manager of a company in trouble has an incentive not to disclose private information.

In principle, an analyst would like to observe the entire event horizon to discount all possible scenarios and understand the value of a company and its investment projects. However, this is not possible because it is the typical double problem of a financial institution facing uncertainty in the presence of information asymmetries. On the one hand, the company has a clear information advantage in its investment plans. On the other hand, both the financial institution and the company have a limited ability to predict future economic shocks, which can have either positive or negative effects.

In their seminal works \cite{altman1968financial} and \cite{ohlson1980financial} apply standard econometric techniques---i.e., multiple discriminant analysis (MDA) and logistic regression---to assess the probability of firm bankruptcy. Following these contributions and the Basel Accord II in 2004, default forecasts are based on standard reduced-form linear regression approaches. However, these approaches may fail because their limited complexity precludes nonlinear interactions among predictors, while their ability to handle large sets of predictors is limited due to potential multicollinearity problems.

Machine learning algorithms compensate for these shortcomings by providing flexible models that allow nonlinear interactions in the space of predictors and the inclusion of a large number of predictors without the need to invert the covariance matrix of the predictors, thereby circumventing multicollinearity \citep{linn2019estimating}. In addition, machine learning models are directly optimized to perform the prediction task, resulting in better prediction performance in many complex situations.

The data science literature has already developed exercises to predict corporate failures using financial accounts, but without a clear economic and financial framework. In this context, \citet{bargagli2021supervised} provides a comprehensive review of the recent literature on the use of machine learning for the analysis firm dynamics. The authors highlight that the majority of work dealing with the prediction of bankruptcy or financial distress uses either decision tree-based techniques \cite[see, e.g., ][]{behr2017default, linn2019estimating, moscatelli2019corporate, deyou2020, davies2023predicting, incerti2022two} or neural network-based methods \cite[see, e.g.,][]{alaka2018systematic, bredart2014bankruptcy, hosaka2019bankruptcy, sun2011dynamic, tsai2008using, tsai2014comparative, wang2014improved, lee1996hybrid, udo1993neural}.

\begin{figure}
	\centering
	\caption{Fictional distribution of failure's probability}
	\resizebox{0.7\textwidth}{!}{%
	\label{fig:distress}
	\includegraphics[width=1\linewidth]{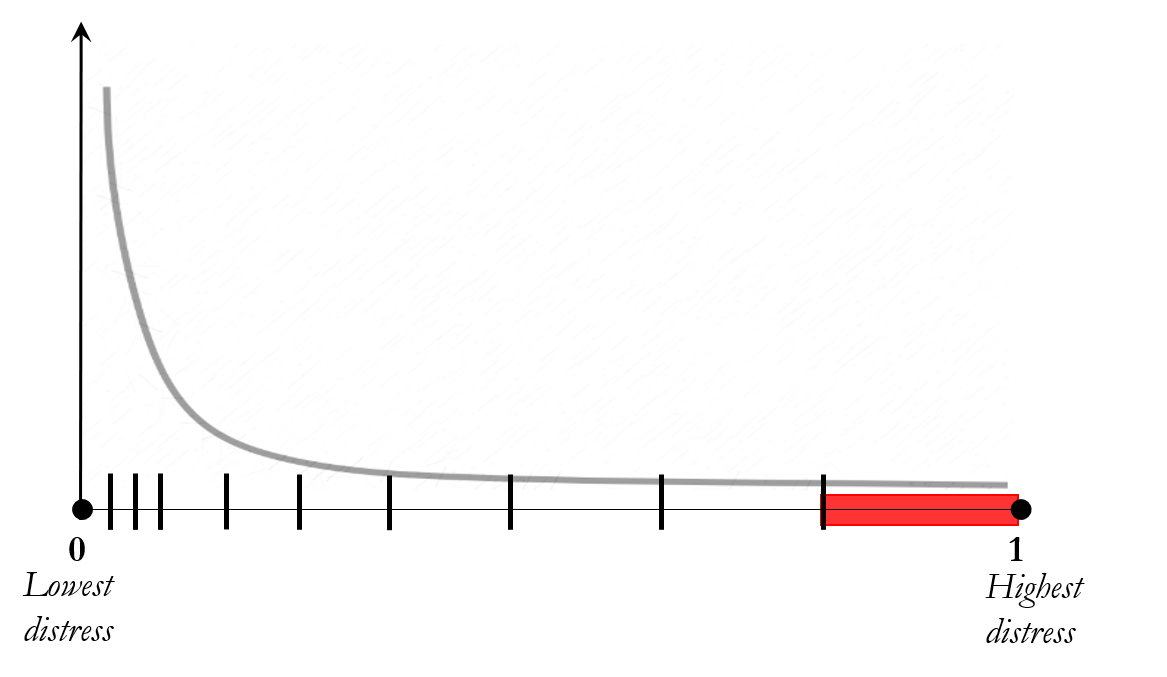}}%
\end{figure}

With this in mind, we propose a machine learning procedure that uses past information about previously failed firms to estimate the probability that another firm in similar condition will go bankrupt. The broader the variety of past experiences on which we can draw, the more accurate the prediction about the health - or lack thereof - of a company \citep{kleinberg2015prediction}. Ultimately, our perspective is on a firm's (lack of) resilience, using potentially any observable data that might hold information about the firm's viability. In the end, we obtain a probabilistic measure at the firm level, ranging from 0 to 1, which tells us the probability that a firm will exit the market in the next period, given that other firms in a similar situation have done so. As shown in Figure \ref{fig:distress}, we can assess the distance of each company from the highest financial distress.

Let us consider a generic predictive model in the form:
\begin{equation} \label{form:pp}
    {f}(\bX_{i,t-1})={Pr}(Y_{i,t}=1 \:|\: \bX_{i,t-1}=x)
\end{equation}
where $Y_{i,t}$ is the binary realization of the outcome at time $t$ that takes the value 1 if the $i$th firm exits the market and takes the value zero otherwise, while $\bX_{i,t-1}$ is the $P$-dimensional vector of firm-level predictors in the previous time period, where $P$ is the number of predictors included in the model. The functional form linking the predictors to the outcomes is determined by the generic supervised machine learning procedure used to predict out-of-sample information. In short, the generic algorithm chooses the best in-sample loss-minimizing function in the form:
\begin{equation} 
\argmin \sum_{i=1}^{N} L(f(x_{i,t-1}), y_{i,t}) \: \: \: \:  over  \: \:  \: \: f(\cdot) \in F \: \: \: \:  \: \: \: \: s. \: t.\: \:  \: \: \: \:  \: \: R\big(f(\cdot)\big) \leq c
\end{equation}
where $F$ is a function class from where to pick $f(\cdot)$, and $R\big(f(\cdot)\big)$ is the generic regularizer that summarizes the complexity of $f(\cdot)$ \cite[see also][]{mullainathan2017machine}. In our case, the function $f(\cdot)$ is an element from the family of classification trees or a combination of them. The set of regularizers, $R$'s, will change according to the standards adopted by each method. Ultimately, each algorithm takes a loss function $L({f}(x_i), y_i)$ as input and searches for the function that minimizes the prediction losses.

\review{To determine the region of highest financial distress, we show in the following analysis that a cutoff of 0.9 minimizes the combination of false-positive (false prediction of firm failures) and false-negative (false prediction of active firms) rates. This empirical result supports the choice of the highest decile of the predicted risk distribution as the optimal threshold for identifying cases of critical financial distress. We will also prove that using the highest decile to identify zombie firms is successful, as this is the segment where prediction accuracy is highest, and after which the probability of transitioning back to lower levels of financial distress is minimal.} 

\bigskip

We would like to emphasize that we are not interested in identifying the causes of firm failure, since we are dealing with a pure prediction problem,--- i.e., the probability of the failure of firm $i$ at time $t$. Nevertheless, we perform variable selection to evaluate the contribution of the predictors to the estimated risk. We show how the predictors of failure can change over time, under the circumstances in which we make the predictions, each time there is an update with new out-of-sample information.

\subsection{Decision tree learning for failure prediction} \label{subsec: warning}

Consistent with the literature on bankruptcy and firm exit predictions presented above, we derive---given new out-of-sample information--- a prediction for the failure of each firm based on its current financial accounts, both for established firms that have operated in previous periods and for firms entering the market for the first time. From another perspective, we interpret this probability range as the degree of risk of an investor who has no information other than that contained in the current financial accounts.

\begin{figure}
    \centering
    \caption{An example of binary tree}
    \captionsetup{font=footnotesize}
    \begin{subfigure}[b]{0.48\textwidth}
        \centering
		\begin{tikzpicture}[level distance=80pt, sibling distance=50pt, edge from parent path={(\tikzparentnode) -- (\tikzchildnode)}]
		\tikzset{every tree node/.style={align=center}}
		\Tree [.\node[rectangle,draw]{$x_1<0.6$}; \edge node[auto=right,pos=.6]{No}; \node[circle,draw]{$l_1$}; \edge node[auto=left,pos=.6]{Yes};[.\node[rectangle,draw]{$x_2 > 0.2$}; \edge node[auto=right,pos=.6]{No}; \node[circle,draw]{$l_2$}; \edge node[auto=left,pos=.6]{Yes}; \node[circle,draw]{$l_3$};  ]]
		\end{tikzpicture}
  \caption{Binary tree}
    \end{subfigure}
    \begin{subfigure}[b]{0.48\textwidth}
    \includegraphics[width=\textwidth]{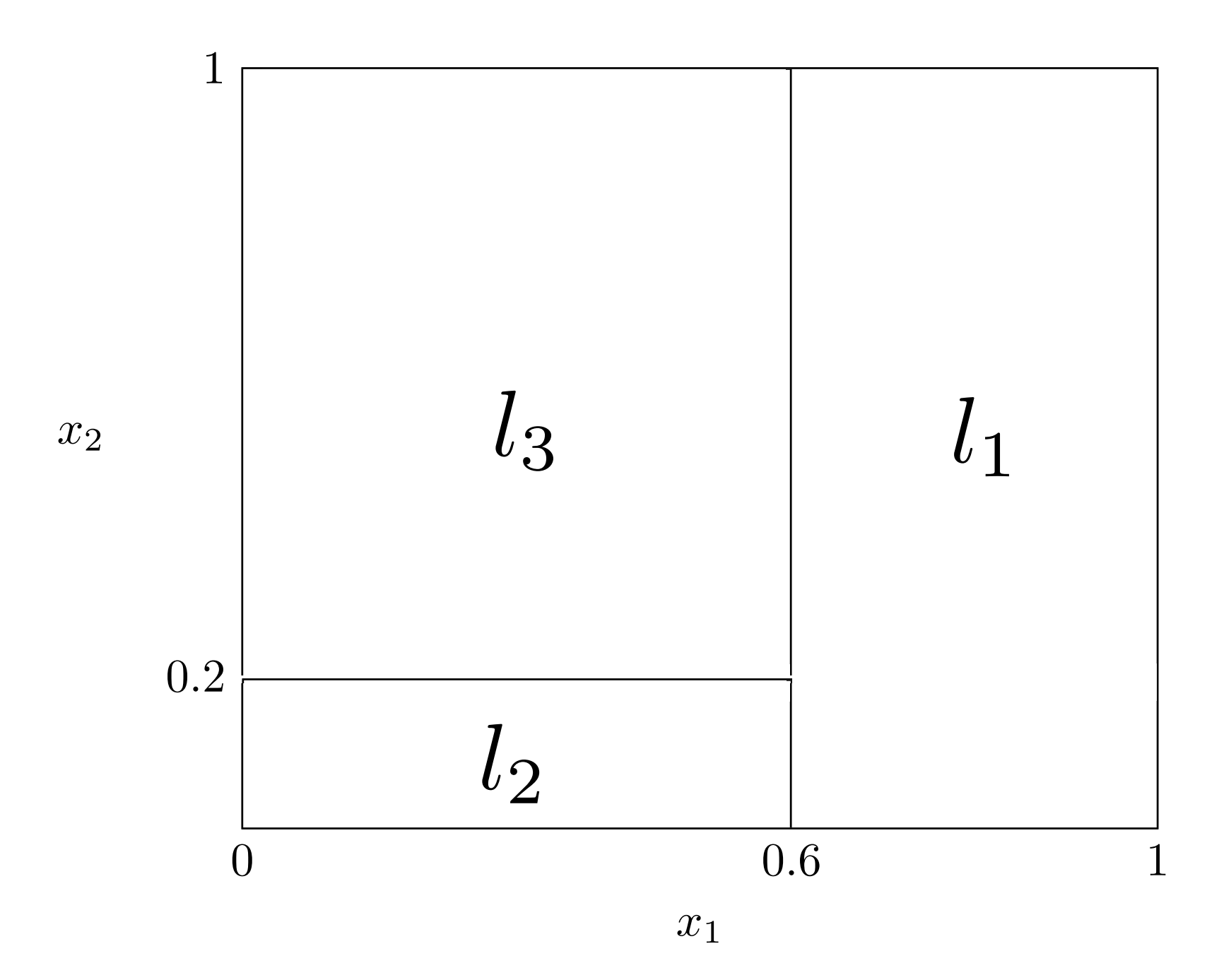}
 \caption{Feature space}
    \end{subfigure}
    \singlespacing
    \caption*{{In Figure (3a), the internal nodes are labeled by their splitting rules and the terminal nodes by the corresponding parameters $l_i$. Figure (3b) shows the corresponding partition of the feature space.}}
    \label{fig:CART}
\end{figure}

\review{Our study introduces an innovative approach to failure prediction by demonstrating the remarkable effectiveness of tree-based machine learning models, especially in the presence of missing data. While neural networks are widely considered the best option for prediction tasks with homogeneous data such as images and text, they often fail for ``problems with heterogeneous features, noisy data, and complex dependencies" \citep{prokhorenkova2018catboost}. Decision tree-based algorithms are considered suitable tools in these cases due to their flexibility and high performance. The Classification and Regression Tree (CART) algorithm, first introduced by \cite{friedman1984classification}, is a widely used decision tree algorithm that constructs binary trees where each node is divided into only two branches. Figure \ref{fig:CART} shows how binary partitioning works in practice, using a simple example with only two predictors. Ensemble methods have been developed to improve the stability of decision tree estimates, which combine multiple weak learners into one strong learner. Bagging and boosting are two popular methods for this purpose \citep{breiman1996bagging, freund1996experiments}. In line with recent developments in the literature, we perform a comparative evaluation of firm failure prediction using two state-of-the-art approaches: XGBoost \citep{chen2016xgboost} and BART-MIA \citep{chipman2010bart}. Both statistical models are specifically designed to deal with missing values, which makes them ideal for dealing with severe cases of missing financial data in corporate accounts. As illustrated in Section \ref{sec:data}, the absence of financial data does not follow random patterns, or at least does not follow the failure of a company completely randomly. When a company is financially distressed and/or smaller, it is more likely that data is missing from some accounts. Simply discarding the missing observations would introduce selection bias and exclude certain categories of firms with a higher probability of failure. Crucially, both XGBoost and BART-MIA include patterns of undisclosed accounts as a feature of the model.}

\review{XGBoost is an ensemble method based on decision trees with a gradient boosting framework. It has proven to be highly competitive and reliable on various prediction tasks, mainly due to its efficient use of computation time and memory resources \citep{gumus2017crude, abbasi2019short, li2020diabetes}. It is equipped with various optimization techniques and tools, such as a percentile-based split-finding algorithm, parallelized tree building, a depth-first approach to tree pruning, and efficient handling of missing values through \textit{default directions}. In addition, it uses regularization to prevent overfitting \citep{bentejac2021comparative}. 

BART-MIA is a robust Bayesian ensemble of trees methodology that combines the traditional BART \citep{chipman2010bart} with a variant called MIA \citep{twala2008good}. This variant was specifically designed to deal with MNAR (Missing Not At Random) patterns in the data. BART is a well-established method that has shown excellent performance on various prediction tasks \citep{murray2017log, linero2018bayesian1, linero2018bayesian2, hernandez2018bayesian}. This can be attributed to two key factors: first, BART contains a noise component that mitigates the overfitting problem common to random forest methods, and second, it has shown consistently strong performance on standard model specifications, avoiding time-consuming hyperparameter tuning procedures. This property is particularly important because it reduces the researcher's dependence on parameter choice and minimizes the computational time and cost associated with cross-validation. See the appendix \ref{appendix:methods} for the technical details of the two models.}

\subsection{Validation against other machine learning techniques} \label{sec:result}

\review{We compare the predictive performance of our methods that account for missing values, XGBoost and BART-MIA}, with the Conditional Inference Tree \citep{hothorn2006unbiased}, the Random Forest \citep{breiman2001random}, and the Super Learner \citep{van2007super}. The Conditional Inference Tree we use is a simple variant of the Classification and Regression Tree (CART) algorithm \citep{friedman1984classification}, based on a significance testing procedure that avoids bias towards variables with many possible splits \cite[see][]{oden1975arguments, loh2002regression, hothorn2006unbiased}. The Random Forest is an ensemble method that combines different trees to obtain stronger predictive power. Each tree is created by randomly selecting different variables from all possible predictors and randomly selecting a subset of the total number of observations \cite[see also][]{breiman2001random}. The Super Learner \citep{van2007super} is based on a weighted combination of other algorithms. \review{We build it as a convex combination of the following models: Logistic regression, CART, Random Forest, BART, and XGBoost.

To evaluate the performance of the models, we use a standard five-fold cross-validation, where the dataset is divided into five groups and the model is trained on four of them, while the fifth group is used as a validation set, iteratively repeating this process. In this way, we can comprehensively evaluate the predictive performance of the model over the entire time series. 

To ensure comparability with the other models that do not consider missing values, we perform a Complete-Case analysis. The latter is a widely used approach for dealing with missing values in models that are not designed to do so. In this approach, all observations with at least one missing value in the predictors are discarded. Although XGBoost and the BART-MIA models could use the entire dataset without excluding missing observations, we perform a Missing-Aware analysis to ensure fair performance evaluation of all models: we train and test XGBoost and BART-MIA using five-fold cross-validation with identical dimensions as the original folds but allow missing observations to be included in the sample. For clarity, we use the terms Missing-Aware XGBoost (MA-XGBoost) and BART-MIA for the models trained on observations with missing values. In this case, MA-XGBoost uses \textit{default directions} to deal with missing values, while BART-MIA uses the MIA procedure. On the other hand, we will simply refer to the models trained on Complete-Case data as XGBoost and BART.}

\begin{table}[H]
\caption{\review{Models’ horse race: performance measures}}
\label{tab:performance_measures}
\captionsetup{font=footnotesize}
\begin{subtable}[h]{\textwidth}
\caption{Complete-Case analysis}
\centering
    \begin{tabular}{lcccccc}
\hline 
\multicolumn{1}{c}{Method}  &  AUC  &  PR  &  F1-Score  &  BACC  &  $R^2$  &  Time \\ \hline \hline  
\textit{Logit} &     0.8966   &   0.4542   &   0.1833   &   0.7504   &   0.2658   & 9.13 \\ 
\textit{Ctree}  &   0.8957  &   0.4444   &   0.1987   &   0.7668   &   0.2640  &   572.46    \\ 
\textit{Random Forest}  &   0.9117   &   0.5233   &   0.1907   &   0.7595   &   0.3135  &   261.62    \\ 
\textit{XGBoost} &    0.9140   &   0.5170   &   0.1833   &   0.7504   &   0.3126 &   43.66    \\ 
\textit{BART}   &    0.9185  &   0.5221   &   0.1843   &   0.7533   &   0.3179   &   1249.05    \\ 
\textit{Super Learner} &   0.9231   &   0.5464   &   0.1844   &   0.7535   &   0.3373   &   4147.87   \\ \hline 
\multicolumn{1}{c}{}
\end{tabular}
\label{tab:cc_analysis}
\end{subtable}
\hfill
\begin{subtable}[h]{\textwidth}
\caption{Missing-Aware analysis}
\centering
    \begin{tabular}{lcccccc}
\hline 
\multicolumn{1}{p{\widthof{Random Forest}}}{Method} &  AUC  &  PR  &  F1-Score  &  BACC  &  $R^2$  &  Time \\ \hline \hline
\textit{MA-XGBoost} &  0.9685     &     0.7591  &  0.2070     &     0.7646     &   0.5243  &  24.70    \\
\textit{BART-MIA} &   0.9681   &    0.7516  &  0.2092 &    0.7676   &    0.5178    &    1126.88    \\ \hline
\end{tabular}
\label{tab:missing_aware_analysis}
\end{subtable}
\singlespacing
\caption*{\review{All algorithms are trained with five-fold cross-validation. The training and test sets include 95,970 and 19,194 observations in each iteration, respectively. All metrics correspond to the five-fold average. Time indicates the average seconds required to train the model in each fold.}}
\end{table}

\begin{figure}[H]
\centering
  \caption{\review{Goodness-of-Fit Scores}}
  \label{fig:auc_and_pr}
  \captionsetup{font=footnotesize}
 \begin{minipage}{.5\textwidth}
  \centering
  \label{fig:test1}
  \includegraphics[width=1\linewidth]{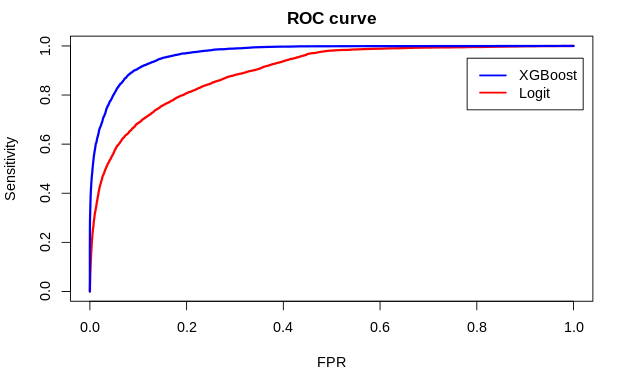}\par
\end{minipage}%
\begin{minipage}{.5\textwidth}
  \centering
  \label{fig:test2}
   \includegraphics[width=0.98\linewidth]{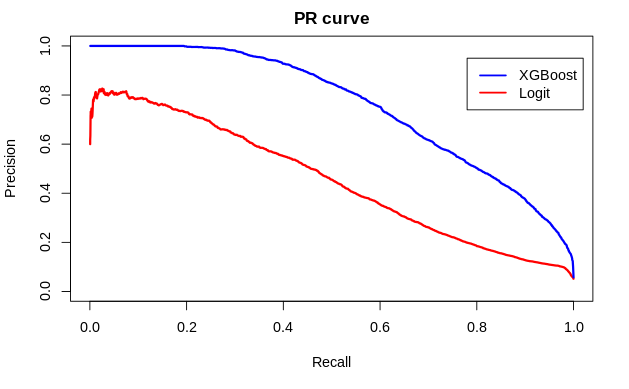}\par
\end{minipage}
\caption*{\review{The ROC and PR curves for the Complete-Case Logit and the MA-XGBoost models. Each plot shows the five-fold cross-validated mean curves, with the mean taken with respect to the ROC and PR curves of each validation set along the cross-validation routine.}}
\end{figure}

Table \ref{tab:performance_measures} shows the results of the horse race of the models. It is clear that the missing-aware models consistently outperform the state-of-the-art methods involved in the Complete-Case analysis. To compare the predictive power of the different methods, we show five different performance measures commonly used for classification problems: the Area Under the receiver operating characteristic Curve (AUC); the area under the Precision-Recall (PR) curve; F1-Score; Balanced ACCuracy (BACC); adjusted $R^2$. Both AUC and PR vary between 0 and 1, with 0 indicating complete misclassification and 1 indicating perfect prediction. The AUC \citep{hanley1982meaning} is a general measure of predictive power that tells us the extent to which we are able to classify failures \textit{vis \`{a} vis} non-failures, hence with an accent on the \textit{false discovery rate} (FDR). PR is particularly useful for our scope because it takes into account both the total proportion of true failures that we are able to predict in the data (i.e., the \textit{sensitivity/recall} of the predictions) and for the proportion of predicted failures that turn out to be true failures (i.e., the \textit{precision} of the predictions). Indeed, assessing \textit{sensitivity/recall} alone could be misleading in a zero-inflation environment such as ours, where the number of non-failures systematically exceeds the number of failures.\footnote{For more details, see \cite{saito2015precision, fawcett2006introduction}.} Figure \ref{fig:auc_and_pr} reports the average ROC and PR curves over the five-folds for the Complete-Case Logit and  \review{MA-XGBoost}.
The F1 score \citep{van1979information} and the BACC \citep{brodersen2010balanced} are used for cases of unbalanced data. The former is based on a harmonic mean of \textit{precision} and \textit{recall}, while the latter is a simple average between the rate of true positives and the rate of true negatives from our predictions.

\review{In the Complete-Case analysis, the best performing model is the Super Learner, which, however, stops at $0.9231$ and $0.5463$ in terms of AUC and PR, respectively.\footnote{It is noteworthy that the Super Learner performs better than the Random Forest in terms of accuracy, but not in terms of F1 score. The reason is that the Super Learner algorithm \citep{van2007super} is optimized to find a convex combination of algorithms that minimizes the accuracy of the ensemble method. The latter strategy is not optimal in our case because the dataset is unbalanced.}
The results of the Missing-Aware analysis differ significantly from those of the Complete-Case ones. MA-XGBoost shows a PR of $0.7591$ and an AUC of $0.9685$, followed by BART-MIA, whose performances are almost identical. Interestingly, the most notable improvement comes from higher precision. This is reflected in the larger departure of the Missing-Aware models with respect to the performance measures that use precision (namely, PR and F1 score). This is critical to our objective as it means that Missing-Aware models have better predictive power in identifying companies that will close within a year based on the predictions. However, training the BART-MIA model is computationally intensive and takes 45 times longer than MA-XGBoost. Due to its significantly lower computational cost and higher predictive power, we choose MA-XGboost as the optimal prediction algorithm for the following analysis.

To ensure the robustness of our results, we perform a number of robustness checks in Appendix F. We investigate whether (i) imputing missing observations could improve the performance of predictive models that do not include information on missing attributes; (ii) whether excluding liquidations from the set of firm failures would affect the performance of the model. For (i), we use out-of-range and median imputation methods \citep{josse2019consistency}. We find that the performance of traditional models increases significantly when missing values are considered, especially for tree-based methods. This result is consistent with our observation that missing entries signaled by a constant value either over the entire dataset (out-of-range imputation) or over a specific feature (median imputation) are included in tree splitting. As a result, the model can now account for missingness in a similar way to the standard directions and MIA techniques, leading to comparable results. With respect to (ii), we test whether predictive performance depends on the type of failure (e.g., bankruptcy, dissolution, M\&A, and liquidation). We find that excluding the largest group of business failures -- i.e., liquidations -- has little effect on the performance of the models, suggesting that the dependence on the type of failure is negligible.}

\subsection{Validation against proxy models of credit scoring} \label{sec:scores}

So far, we have compared the predictions of failures from different econometric and machine learning techniques and concluded that \review{MA-XGBoost is the best choice in terms of predictive power and computational burden.} Here, we compare our baseline predictions with widely known proxies for firm-level credit scores: the \textit{Z-scores} \citep{altman1968financial} and the \textit{Distance-to-Default} \citep{merton1974pricing}.

Z-scores involve putting a selection of financial ratios (profitability, leverage, liquidity, solvency, and volumes of activity) into an equation with some weights to proxy their relative importance. The weights are taken from the literature and from previous scholarly estimates of the relative importance of these indicators in assessing a firm's distress. In this way, a threshold is obtained, the crossing of which indicates a high probability of future bankruptcy.
Unlike Z-scores, the Distance-to-Default (DtD) by \cite{merton1974pricing} focuses specifically on a firm's ability to meet its financial obligations. The original intuition is that a firm's equity can be modeled as a call option on its assets. Thus, to build such a model, one must combine firm-level accounts (firm assets, debt, market value) and information from financial markets (risk-free interest rate, standard deviation of stock returns). Finally, one puts the variables into an equation that gives the value of a theoretically fair call option.\footnote{Following the insights of the distance-to-default model, \cite{black1973pricing} developed their widely known model based on the observation that one can eliminate a systemic risk component by hedging an option.}

\review{We evaluate the predictive power of both Z-scores and Distance-to-Default (DtD). The Precision and False Discovery Rate (FDR) are given in Table \ref{tab:indicators_vs_bart}. Similar to our previous analysis, we performed five-fold cross-validation. We start with the non-missing observations in Z-scores and DtD and randomly split them into five folds. Based on the Z-scores and DtD measures, distressed firms have lower scores. Therefore, we used the first ten percentiles of the in-sample scores as cutoffs for classifying the out-of-sample observations. We found that the DtD predictions have higher precision (0.3314 vs. 0.2239) and lower False Discovery Rate (0.6686 vs. 0.7761) than the Z-scores. We then train a MA-XGBoost model with the same cross-validation routine and use the in-sample percentiles as cutoffs to classify the out-of-sample observations. The MA-XGBoost model considers the highest risk firms to be in the right tail, unlike Z-scores and DtD. Therefore, the first column of Table \ref{tab:indicators_vs_bart} represents the bottom ten percentiles (1-10) for DtD and Z-scores, while for MA-XGBoost the top ten percentiles (99-90) symmetrically. It is clear that MA-XGBoost outperforms both DtD and Z-scores in all percentiles.}

\begin{table}[H]
\centering
\caption{\review{Goodness-of-fit: Distance-to-Default (DtD), Z-scores and MA-XGBoost}}
\captionsetup{font=footnotesize}
\label{tab:indicators_vs_bart}
\centering
\resizebox{0.7\textwidth}{!}{%
\begin{tabular}{cccccccc}
\hline
& \multicolumn{2}{c}{DtD} & \multicolumn{2}{c}{Z-Scores} & \multicolumn{2}{c}{MA-XGBoost} \\ 
\cmidrule{2-7}
Percentile  & Precision & FDR & Precision & FDR & Precision  & FDR  \\ \midrule
        1 & 0.3314 & 0.6686 & 0.2239 & 0.7761 & 0.9850 & 0.0150 \\  
        2 & 0.3314 & 0.6686 & 0.2102 & 0.7898 & 0.9054 & 0.0946 \\  
        3 & 0.3314 & 0.6686 & 0.2070 & 0.7930 & 0.8316 & 0.1684 \\  
        4 & 0.3314 & 0.6686 & 0.1986 & 0.8014 & 0.7517 & 0.2483 \\  
        5 & 0.3020 & 0.6980 & 0.1937 & 0.8063 & 0.6715 & 0.3285 \\  
        6 & 0.2723 & 0.7277 & 0.1882 & 0.8118 & 0.6037 & 0.3963 \\ 
        7 & 0.2497 & 0.7503 & 0.1875 & 0.8125 & 0.5447 & 0.4553 \\  
        8 & 0.2334 & 0.7666 & 0.1831 & 0.8169 & 0.5018 & 0.4982 \\  
        9 & 0.2226 & 0.7774 & 0.1769 & 0.8231 & 0.4600 & 0.5400 \\  
        10 & 0.2139 & 0.7861 & 0.1745 & 0.8255 & 0.4312 & 0.5688 \\
\bottomrule
\end{tabular}}
\singlespacing
\caption*{\review{For DtD and Z-Scores, a firm is predicted to fail if its score falls below a certain threshold. The thresholds for classification are determined by the first ten percentiles of their in-sample score distribution. In contrast, MA-XGBoost predicts failure when the score is above a certain threshold, and classification cutoffs are determined by the top ten percentiles of the own in-sample score distribution.}}
\end{table}

\subsection{Unboxing the black box: Shapley values for predictors} \label{sec:indicators}

In the previous sections, we validated our proposed  \review{MA-XGBoost} model against state-of-the-art financial indicators, traditional econometric models, and widely used machine learning techniques. Despite its excellent predictive performance, the nonlinear relationships learned by \review{MA-XGBoost} are not directly observable, leading to the recurring criticism that machine learning models are black boxes despite their improved accuracy. However, there are a number of tools to shed light on complex nonlinear relationships in machine learning predictions to make models interpretable \citep{lundberg2020local}.\footnote{Interpretability is a non-mathematical concept, but is often defined as the degree to which a human can understand the cause of a decision or consistently predict the outcomes of the model \cite[see, e.g.,][]{kim2016examples, miller2018explanation, stoffi2021assessing, lee2020causal, bargagli2022heterogeneous, bargagli2020essays, bargagli2020causal}.} Popular methods for assessing predictor importance include permutation importance or Gini importance for tree-based models \citep{breiman2001random}, LIME \citep{ribeiro2016model}, DeepLIFT \citep{shrikumar2017learning}, and Shapley values \citep{strumbelj2010efficient}.

We chose to use Shapley values because they are not only able to reveal the complex patterns linking the predictors and the outcome, but also guarantee a number of desirable properties: \textit{efficiency, null payoffs, symmetry, monotonicity} and \textit{linearity} \citep{rozemberczki2022shapley}.
In the case of a general ML model, these properties state that: the importance of individual variables must add up to the goodness-of-fit of the model trained with the entire set of variables (\textit{efficiency})\footnote{This property is crucial because it allows quantifying the contribution of each variable to the overall performance of the model.}; a variable that does not improve the goodness-of-fit of the model is assigned a contribution of zero (\textit{null payoff}); two variables that make the same marginal contribution to the global goodness-of-fit have the same importance (\textit{symmetry}); if variable A consistently contributes more to the global goodness-of-fit of the model than variable B, the Shapley value of A must be higher than that of B (\textit{monotonicity}); for two subsets of the same data set, the global Shapley value of a variable is equal to the sum of the Shapley values calculated separately for the same variable in the two subsets (\textit{linearity}).

An intuitive way to understand how Shapley values for variable importance work is to include each variable in the predictive model in random order. All variables in the model contribute to the final predictions (and thus to the goodness-of-fit of the model). The Shapley value of a variable is the average change in prediction (and goodness-of-fit) experienced by the set of variables already in the model when that variable is included \citep{molnar2020interpretable}. From a mathematical perspective, assume that $S$ is a $q$-dimensional subset of variables, $m$ is a generic variable in $S$ ($m \subset S$), and $v(T)$ is a generic value function that takes in the subset $S$ and returns real-valued payoff of the model (e.g., the goodness-of-fit) created using $S$ or subsets thereof. Then the Shapley value $\phi_m(v)$ for a generic variable $m$ is:
\begin{equation}\label{eq:shap}
    \phi_m(v) = \frac{1}{q} \sum_{S \subseteq \{1, ..., q \} \setminus \{m\} } [v(S \cup \{m\}) - v(S)] \frac{|S|!(q - |S| -1)!}{q!}.
\end{equation}
Using \eqref{eq:shap}, we see how the Shapley value is computed by calculating a weighted average gain in payoff (read: gain in goodness-of-fit) that the variable $m$ yields when included in all subsets of variables that exclude $m$.

\begin{figure}[H]
\centering
\caption{\review{Shapley values for the variables in the predictive model}}
\label{fig:shap_1}
\includegraphics[width=0.9\textwidth]{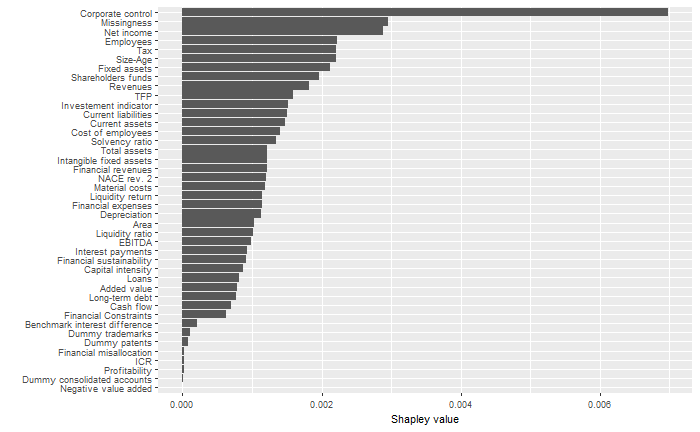}
\end{figure}

\begin{figure}[H]
\centering
\caption{\review{Shapley values for the groups of variables in the predictive model}}
\label{fig:shap_2}
\includegraphics[width=0.9\textwidth]{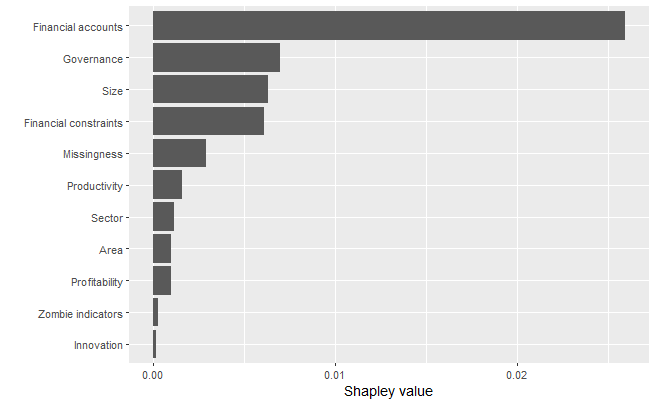}
\end{figure}

The results of our Shapley value analyses are shown in Figures \ref{fig:shap_1} and \ref{fig:shap_2}. Figure \ref{fig:shap_1} reports the results of the average Shapley Values over the time frame of our analysis for each variable used as input. Figure \ref{fig:shap_2} shows the same value, but this time all variables are aggregated into some economically relevant groups.\footnote{A full description of how variables are aggregated into groups is described in Appendix Table A1. Governance variables are indicated by (G), financial constraints by (FC), financial accounts by (FA), zombie indicators by (ZI), area by (A), innovation by (I), productivity by (PDC), sector by (SE), profitability by (PFT), size by (SI).} 

\review{First, we note that no single financial indicator has better predictive power than the aggregate of indicators. On their own, most indicators have relatively low predictive power. The financial indicators with higher Shapley Values -- i.e., that contribute most to the model's predictions -- are the corporate control indicator (\textit{Corporate control}), the missingness in the financial accounts (\textit{Missingness}), some profit and loss characteristics (\textit{Net income, Taxes, Revenues}), the size and age indicators (\textit{Employees, Size-age}), the resources invested in the company (\textit{Shareholders funds}) and the long-term physical assets (\textit{Fixed assets}).\footnote{See Table \ref{tab:table2} for a full description of the predictors and their construction.} When analyzing the groups of indicators, it appears that original financial accounts with no further elaboration carry the greatest explanatory power. This result is reasonable considering that most predictors fall into this category. In addition, the governance, size, and financial constraint indicators also have relatively high predictive power. Although they compete with groups consisting of multiple predictors, missingness remains an important factor. On the other hand, firms' productivity, sector, and geographic location of firms also have high predictive power, but are of secondary importance compared to the aforementioned indicators.

Finally, to check for robustness, we conduct a logit-LASSO analysis to evaluate, using a shrinkage method, the most frequently selected variables that predict a firm's distress. We find substantial overlap, with corporate control, profit and loss characteristics, and firm size-age indicators among the most important predictors. The full analysis can be found in the Appendix \ref{appendix:lasso}.\\

We conclude this Section with a discussion of the possibility that firms manipulate financial accounts and the predictive power of missing values in this case. We argue that there are two reasons for the occurrence of missing values in firms' financial accounts:
\begin{enumerate}
 \item Smaller firms are often exempt from reporting complete information to national registries if they fall below certain size thresholds because it may be too costly for them to implement complex accounting procedures.\footnote{In footnote 7 of the manuscript, we indicate the combination of size thresholds that the Italian law provides.}
 \item Some financial accounts are optional each year or may be incomplete (e.g., number of employees, management of inventories, cash prospects, etc.).
\end{enumerate}
In any case, there is room for manipulation by firms that know that banks and policymakers do not use random missing values to predict their financial distress. They have two choices: they can provide truthful information and thereby disclose their true financial distress, or they can provide false information and thereby potentially commit accounting fraud. In the first case, the missing values are no longer meaningful, but our algorithm can still rely on the disclosed financial accounts for prediction. In the second case, we would have false negatives, i.e., companies that are predicted to be financially viable but are not. More generally, we conclude that a limitation of our approach is that we assume that companies do not report false information; otherwise, we would have statistical noise that reduces prediction accuracy.}

\section{A case for \textit{zombie firms}\label{sec: zombies}}


\review{In our analysis, we showed how we can use machine learning to predict firms' failures.} But how do we spot a \textit{zombie firm}? There is no consensus on the exact meaning of the \textit{zombie} status, other than its suggestive power. To date, scholars have merely adopted various thresholds based on a proxy assessment of one or more available financial indicators in the absence of more precise theoretical guidance. Ideally, a company's competitiveness and financial constraints should be viewed from a dynamic perspective that considers the entire horizon of future events. Considering all future threats and opportunities and their impact on profit and net cash flow at the firm level could be the theoretical equivalent of ``\textit{zombie firms}". Without the latter, empirical identification is left to the creativity of academics and practitioners. \cite{caballero2008zombie} define \textit{zombies} as firms that receive subsidies in the form of bank loans after observing how interest payments compare to an estimated benchmark of debt structure and market interest rates.
\cite{mcgowan2018walking} assume that \textit{zombies} are old firms that have persistent problems meeting their interest payments, although they focus a policy discussion on the macroeconomic impact of low-productivity firms. \cite{bankofkorea2013} explicitly examines when the interest coverage ratio (ICR) is below one over three years. \cite{bankofengland2013} disregards financial management and considers firms that have both negative profits and negative value added, thus focusing on a firm's core activity.

In our view, the direction of the work so far is clear: scholars and practitioners want to infer companies' future viability from their current financial accounts. If they do not appear to be in good shape, it is likely that the company will be in trouble in the near future. From this perspective, the empirical classification of \textit{zombie firms} is a perfect case study for applying machine learning techniques to firm-level data. It is a call to use in-sample information to predict an out-of-sample event.

Strengthened by the previous intuition, we propose an identification of \textit{zombies}, starting from the predictions of failure made in Section \ref{sec:strategy}. We propose to classify as \textit{zombies} those firms that are at the right end of the risk distribution and for which the chances of recovering from financial distress are minimal for at least three years.

In notation, we first consider the deciles along the predictions ${f}(\bX_{i,t-1})$, where each $q_{j,t}$ is the threshold for the $j$th decile of the probability of default at time $t$:
\begin{equation}
		Q_{j,t}=\begin{cases}
		1 &	 \text{if ${f}(\bX_{i,t-1}) \geq q_{j,t}  \:\cap\: Y_{i,t} \neq 1$}, \nonumber \\
		0 &	  \text{otherwise}.
		\end{cases}
\end{equation}
where $Y_{i,t}\neq 1$ indicates that the $i$th firm did not fail yet at time $t$.

\vspace{0.5cm}

\begin{figure}[H]
\centering
\caption{\review{Transitions after predictions of a \textit{zombie} status}}
\captionsetup{font=footnotesize}
\label{fig:zombie_status}
\resizebox{0.8\textwidth}{!}{%
\includegraphics[width=1\textwidth]{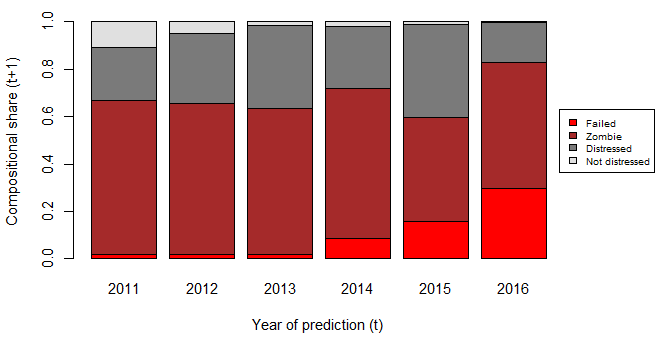}}
\caption*{\review{The bars of the diagram show the transition of \textit{zombie} firms in the years following the prediction: i) to failure (light red); ii) to remaining in a \textit{zombie} status (dark red); iii) to relatively lower distress, i.e. between the 6th and 9th deciles (dark gray); iv) to a range of no distress, i.e., below the 6th decile (light grey). Note that we cannot report 2017 because we cannot compare it to actual observations in subsequent years.}}
\end{figure}

\begin{table}[H]
\centering
\caption{\review{Transitions across deciles of risk}}\label{tab:transition}
\resizebox{1\textwidth}{!}{%
\begin{tabular}{ccccccc}
\toprule$t\:/\:t+1$
 & 9th decile $t+1$ & 8th decile $t+1$ & 7th decile $t+1$ & 6th decile $t+1$ & Below 6th decile $t+1$ & Total $t+1$\\ 
\midrule
9th decile $t$ & 0.46 & 0.22 & 0.12 & 0.08 & 0.12 & 1.00 \\ 
8th decile $t$ & 0.22 & 0.22 & 0.17 & 0.13 & 0.26 & 1.00\\ 
7th decile $t$ & 0.11 & 0.16 & 0.19 & 0.15 & 0.39 & 1.00\\ 
6th decile $t$ & 0.07 & 0.11 & 0.15 & 0.18 & 0.49 & 1.00\\ 
Below 6th decile $t$ & 0.03 & 0.04 & 0.06 & 0.09 & 0.78 & 1.00\\ 
\bottomrule
\end{tabular}}
\end{table}
\vspace{0.5cm}
\review{
In Table \ref{tab:transition}, we report a transition matrix for the firms that did not fail, based on elaborations over the entire period of analyses in 2008-2017. We observe that a significant share of firms that our predictions locate beyond the 9th decile in a representative year $t$ do not have a high chance to improve in $t+1$. In fact, most of them (46\%) remain stuck in the same highest-risk category, and only 12\% are able to recover and reach a more reasonable level of financial distress, i.e. below the 6th decile. Interestingly, the 9th decile is quite difficult to reach from the bottom of the distribution, as only 11\%, 7\%, and 3\% of firms from lower deciles, respectively, are observed to transit to a situation of highest distress. Moreover, about 78\% of companies stay below the 6th decile in the entire analysis period.  

In general, we can say that, according to the information we have observed, a viable firm does not easily shift into financial distress, but if it does, it is difficult to recover from it. With this in mind, it makes sense to set an appropriate threshold that realistically reflects the most difficult situations in business life. We obtain this threshold by determining the cutoff that minimizes the combination of false positive and negative rates. We accomplish this task by maximizing the BACC, since it corresponds to a convex combination of the true positive and negative rates, which in turn are complementary to the false positive and negative rates. Moreover, the concept of zombie firms is closely related to the concept of false positives, since these firms are expected to fail but remain active. Therefore, focusing on the false positive and negative rates within this framework is natural. According to our results in Figure \ref{fig:bacc}, the BACC peaks at about 0.9.

The latter result provides a first justification for using the 9th decile as a cutoff. Nevertheless, we want to give firms a trial period to discount the break-even strategies of some firms in the short run, e.g., in the case of newly formed firms and start-ups. For all the above reasons, we define \textit{zombie firms} those firms that \textit{persist} at least three years beyond the 9th decile of the risk distribution:

\begin{equation} \label{eq:4}
    \mathbbm{1}(\sum_{t=1}^3  Q_{j,t} =3)
\end{equation}
where the risk distributions are estimated with the MA-XGBoost, as proposed in Section \ref{sec:strategy}.

\begin{figure}[H]
         \caption{\review{BACC at different cutoffs along the distribution of predictions}}
         \centering
         \includegraphics[width=0.6\linewidth]{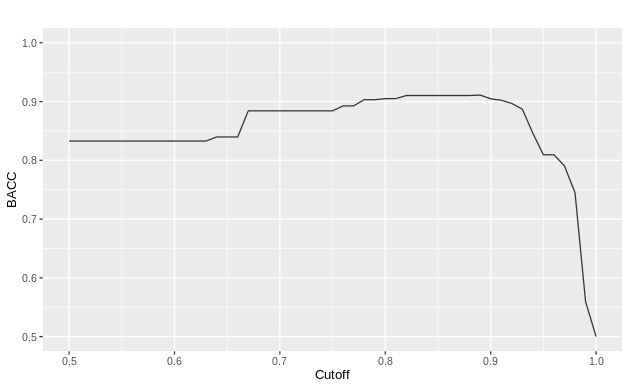}
         \label{fig:bacc}
\end{figure}

Finally, Figure \ref{fig:zombie_status} shows a transition analysis of zombie firms based on our definition from Equation \ref{eq:4}. We calculate the fraction of zombies that either fail, remain in \textit{zombie} status, transition to a lower (but still significant) risk category, or transition to a non-distressed state in the following year. These results suggest that the most common outcome is remaining in zombie status, which applies to nearly 50\% of them over multiple years. In addition, a substantial proportion of zombies transition to a lower-risk category. However, only a small fraction of them completely deviate from their current state, resulting in either failure or absence of distress.}  

\subsection{\textit{Zombies} in Italy\label{sec: zombies in Italy}}

\review{In this section we give some coordinates for the phenomenon of \textit{zombies} in Italy. First, we provide an overview of their evolution along the macroeconomic cycle. Then, we show how they differ from the rest of the viable firms that survive the market. Finally, we provide a comparison across geography and industry taking into account their financial performance and their ability to generate added value.}

\begin{figure}[H]
  \centering
  \caption{\review{\textit{Zombie firms} and the economic cycle}}
  \label{fig:gdp_pdf_9}
  \captionsetup{font=footnotesize}
\resizebox{0.8\textwidth}{!}{%
  \includegraphics[width=1\textwidth]{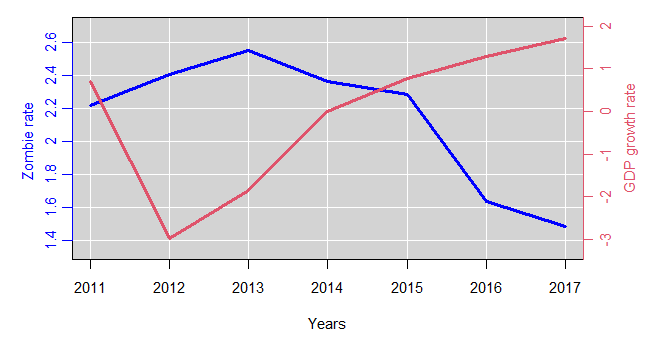}}
\caption*{\review{The share of \textit{zombies} on the left axis is compared with nominal GDP growth rates on the right axis, obtained from the World Bank for the period 2011-2017. \textit{Zombie firms} are firms that are at the right end (9th decile) of the predicted risk distribution for at least three consecutive years.}}
\end{figure}

In Figure \ref{fig:gdp_pdf_9}, we plot the share of \textit{zombies} in our analysis period against the GDP growth rates observed over the same period. Interestingly, we find that a range between \review{$1.48\%$ and $2.55\%$} of manufacturing firms are on the verge of bankruptcy. The share of \textit{zombies} is higher immediately after the financial crisis in 2011 and then decreased from 2013. In fact, the presence of \textit{zombies} seems to be related to the business cycle. The latter is an interesting finding for further analysis but beyond the scope of this paper. We presume it makes sense for \textit{zombie firms} to be countercyclical: many firms can be pushed to the brink of bankruptcy in times of crisis, while a few financially distressed firms can find a way to recover when the recovery begins.

\review{To fully capture their economic importance, we briefly show in Table \ref{tab:prodsize} what distinguishes \textit{zombies} from the other firms. We find that they have lower productivity on average, i.e., 19.7\% and 45\% when measured as TFP and labor productivity, respectively. They are also consistently smaller on average, selling about 200\% less and employing 17\% less than the otherwise representative healthy firm. \review{The latter results are consistent with the idea that these firms drag down aggregate productivity while hindering the redistribution of resources \citep{andrews2017confronting, mcgowan2018walking}.}}

\begin{table}[H]
\centering
\caption{\review{Productivity and size premia for \textit{zombies} vs. healthy firms}}
\label{tab:prodsize}
\captionsetup{font=footnotesize}
    \resizebox{0.8\textwidth}{!}{%
        \begin{tabular}{lcccc}
            \toprule
            Indicator (in logs) & Coeff. & Std. Error & N. obs. & Adj. R squared\\
            \midrule
            Total Factor Productivity & -0.197*** & (.037)  & 600,771  & .967 \\
            Labor Productivity & -0.450***  & (.014)  & 559,315 & .110\\
            Sales & -2.066***  & (.053)  & 1,234,750 & .116 \\
            Employees & -0.170***  & (.039) & 1,119,486 & .157\\
            \bottomrule
        \end{tabular}}
\singlespacing
\caption*{\review{The table shows the coefficients of the linear models for panel data with fixed effects at the region and industry levels. We use pooled OLS estimation with cluster-robust standard errors to account for possible correlations within regions and industries. The dependent variable is a measure of firm productivity or size. The main covariate is an indicator that takes the value of one if the firm is classified as \textit{zombie} in a given year of our sample, and zero otherwise. \textit{Zombie firms} are defined as firms that are at the right end of the predicted risk distribution (above the 9th decile) for three consecutive years.}}
\end{table}

\review{Eventually, we lay out how the segment of \textit{zombies} that we find according to our working definition compares to the segments of firms that:
\begin{enumerate}
 \item have problems in making their interest payments because their Interest Coverage Ratio (ICR);
 \item have problems in their core economic activity because they are destroying value, i.e., their value added is negative.
\end{enumerate}
Interestingly, both interest coverage ratio and negative value added have occasionally been proposed as indicators of \textit{zombieness} in previous literature (see Appendix \ref{sec:appendix_a} for more details on these indicators and their previous use).

In Figure \ref{fig:zombie_geography}, the rays of the radar show for NUTS 2-digit regions the proportions of \textit{zombie} firms that we detect with MA-XGBoost (circles) compared to firms that have an ICR of less than one (triangles in panel a) and firms that experience negative value added (triangles in panel b). We find that there is indeed common support when firms are financially distressed because their ICR is too low, and at the same time they are also classified as \textit{zombies} according to our algorithm. This is evident from a common core region in the radar graph bounded by square nodes. However, the two segments do not coincide. If we consider only the ICR, we may have \textit{zombies} that are obviously not in distress but are classified as such by our algorithm. It is noticeable that this occurs mostly in southern and central Italy. Abruzzo (ITF1), Basilicata (ITF5), Calabria (ITF6), Campania (ITF3), Lazio (ITI4), Molise (ITF2), and Puglia (ITF4) host a relatively large proportion of \textit{zombies} according to MA-XGBoost.\footnote{Not surprisingly, patterns of missing values are also more pronounced in southern and central Italy. Between 2008 and 2016, ICR values are missing for 81.79\% of firms in central Italy and for 76.24\% of firms in southern Italy on average.} In these cases, the cause of distress may lie in other aspects of business activity, as our methodology allows us to use information from a wide range of financial predictors covering different aspects of firms' economic life. 

At the same time, panel (a) of Figure \ref{fig:zombie_geography} also identifies firms that have liquidity problems because their ICR is less than one, but our algorithm does not classify them as \textit{zombies}, possibly because they are still solvent thanks to profitable economic activity that allows them to eventually meet their financial obligations and thus overcome temporary liquidity constraints shortages. This is particularly evident in North and Insular Italy, including Emilia-Romagna (ITH5), Friuli-Venezia Giulia (ITH4), Lombardy (ITC4), Piemonte (ITC1), Sardinia (ITG2), Tuscany (ITI1), Valle d'Aosta (ITC2) and Veneto (ITH3). 

\begin{figure}[H]
\centering
  \caption{\review{\textit{Zombie firms} and geography}}
  \label{fig:zombie_geography}
\captionsetup{font=footnotesize}
 \begin{minipage}{.5\textwidth}
  \centering
  \label{fig:ICR1}
  \includegraphics[width=0.9\linewidth]{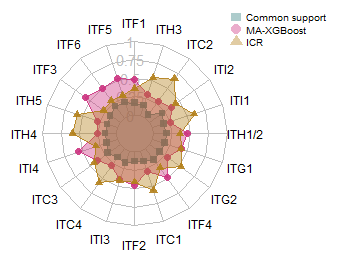}\par
  \subcaption{\textit{Zombies} and interest payments}
\end{minipage}%
\begin{minipage}{.5\textwidth}
  \centering
  \label{fig:NEG_VA1} \includegraphics[width=0.9\linewidth]{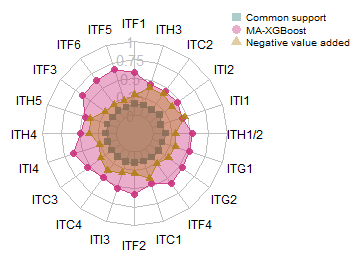}\par
   \subcaption{\textit{Zombies} and value destruction}
\end{minipage}
\singlespacing
\caption*{\review{Note: The rays of the radar show, at the regional level (NUTS 2-digit), the proportion of \textit{zombie} firms versus firms that have an Interest Coverage Ratio (ICR) of less than one (panel a) and versus firms that have negative value added (panel b). The square nodes indicate the common areas where the segments overlap. The circles represent the fraction of \textit{zombies} that we detect using MA-XGBoost. The triangles indicate the fraction of firms identified with $ICR<1$ or negative value added. Along each ray, the values of the squares, circles, and triangles sum to one. Panel (a) and (b) include a total of 30,380 and 24,351 observations, respectively.}}
\end{figure}

On the other hand, in panel (b) of Figure \ref{fig:zombie_geography}, we show that a relatively small segment of firms destroy economic value because what they sell has less value than what they buy as inputs. They are problematic; in fact, most of them are classified  as \textit{zombies}. However, there is a significant portion of \textit{zombies} that show positive value added, but are still struggling financially and are therefore detected as zombies by our algorithm. This scenario extends over the entire country, but peaks in some central and southern regions: Abruzzo (ITF1), Basilicata (ITF5), Calabria (ITF6), Campania (ITF3), Lazio (ITI4), Molise (ITF2) and Puglia (ITF4). The descriptive statistics in Figure \ref{fig:zombie_geography} confirm our intuition that only a comprehensive battery of predictors, including missing values, can provide a clear picture of the relevance of the phenomenon of \textit{zombie} firms. Finally, in the Appendix \ref{sec:appendix_a}, we report in Figure\ref{fig:zombie_industries} and Table\ref{tab:zombies_industries_table} a snapshot of the proportion of \textit{zombies} in different industries (2-digit digits NACE Rev. 2). As largely expected, firms that have problems with interest payments ($ICR 1$) are always a larger segment than firms that we identify as \textit{zombies}. On the other hand, the share of firms with negative value added is generally smaller than the share of \textit{zombie} firms.} 

\section{Conclusions} \label{sec:conclusions}

In this contribution we show how statistical learning can infer non-trivial information on a set of financial indicators, and we successfully classify firms into risk categories after training on past failures. Our preferred algorithm is \review{XGBoost}, which outperforms other well-known econometric and machine learning methods, especially in the presence of missing patterns of financial accounts.

Thanks to our machine learning approach, we can also reduce prediction errors compared to traditional credit scoring tools such as the Z-scores and the Distance-to-Default. Therefore, we propose to classify as \textit{zombie firms} those firms that remain in high-risk status because they are at the right end of the distribution of our predictions for at least three years, beyond the 9th decile of risk, where we find that the chances of recovery to a smaller risk of distress are minimal.

The preceding evidence suggests that identifying \textit{zombies} may be crucial for financial institutions to avoid wasting credit resources on insolvent firms when the latter survive the market only thanks to some adverse selection mechanisms that arise from imperfect financial markets. However, from a more general perspective, we believe that our exercise can be useful to identify a part of an economy that is in trouble. The issue is all the more critical because recent studies discuss how \textit{zombies} can hinder the growth potential of many countries by impeding the redistribution of productive resources in modern economies \citep{andrews2017confronting, banerjee2018rise, andrewspetroulakis}.

Indeed, we find that Italian \textit{zombies} have lower levels of total factor productivity and are mainly found in smaller firms. Interestingly, we note a possible relationship with the business cycle that would be worth investigating further in the future: the share of \textit{zombies} was higher after the two financial crises in 2008 and 2011, but then declined in recent years as the recovery took hold.

It is beyond the scope of this article to make predictions for the post-pandemic scenario. However, we anticipate that the problem of separating companies that can stand on their own two feet from those that hide their inability to pay will become all the more urgent now that policymakers have finally withdrawn financial support programs. The challenge is to avoid further misallocation of resources, which would slow down the much-needed economic recovery.

\newpage
\singlespacing
\setlength\bibsep{3pt}
\bibliographystyle{elsarticle-harv}
\bibliography{sample.bib}

\newpage
\onehalfspacing

\begin{center}
 {\bf \huge Appendix}
\end{center}

\setcounter{table}{0}
\renewcommand{\thetable}{A\arabic{table}}
\onehalfspacing

\appendix 
\counterwithin{figure}{section}
\counterwithin{table}{section}

\section{Tables and graphs} \label{sec:appendix_a}
\begin{table}[H]
    \footnotesize
        \setcounter{mpfootnote}{\value{footnote}}
        \renewcommand{\thempfootnote}{\arabic{mpfootnote}}
        \setlength{\tabcolsep}{4pt} 
        \centering
       \caption{Panel (A): List of predictors for firms' failures.}
       \ContinuedFloat
        \captionsetup{labelformat=empty}
        \label{tab:table2}
        \begin{footnotesize}
        \begin{tabularx}{\textwidth}{>{\raggedright\arraybackslash}p{8cm}>{\raggedright\arraybackslash}p{8cm}}
        \hline\noalign{\smallskip}
        \textbf{Variables} & \textbf{Description}\\ \midrule
        Material Costs (FA), Costs of Employees (FA), Added Value (FA), Taxation, Tax and Pensions' Payables (FA), Revenues (FA), Financial Expenses (FA), Interest Payments (FA), Cash Flow (FA), Fixed Assets (FA), Current Assets (FA), Shareholders' Funds (FA), Retained Earnings (FA), Long-Term Debt (FA), Loans (FA), Current Liabilities (FA), EBITDA (Earnings before interest, Taxation, Depreciation and Amortization) (PFT), Intangible Fixed Assets (FC), Total Assets (SI), Net Income (SI),  Number of Employees (SI) & Original accounts (expressed in euro) \\
        
        \\
        
        Capital Intensity (FA) & Fixed Assets/Number of employees.\\
        
        \\
        
        Corporate Control (G) &  A binary variable equal to one if a firm belongs to a corporate group.\\
        
        \\
        
        Consolidated Accounts (G) & A binary variable equal to one if the firm consolidates accounts of its subsidiaries\\
        
        \\
        
        Number of Patents (I) & The portfolio of patents granted to a firm by patent offices (\textit{Dummy Patents} equal to 0 if the firm issued no patents, and 1 otherwise).\\
        
        \\
        
        Number of Trademarks (I) & The total number of trademarks issued to the firm by national or international trademark offices (\textit{Dummy Trademarks} equal to 0 if the firm issued no trademarks, and 1 otherwise).\\
        
        \\
        
        NACE rev. 2 (SE) & A 4-digit industry affiliation following European classification NACE rev. 2.\\
        
        \\
        
        NUTS 2 regions (A) & The region in which the company is located.\\
        
        \\
        
        TFP (PDT) & It is the Total Factor Productivity of a firm computed as in \cite{ackerberg2015identification}.\\
        \hline
\end{tabularx}
\end{footnotesize}
\end{table}

\begin{table}[H]
\ContinuedFloat
\begin{minipage}[t][\textheight][t]{\linewidth}
    \footnotesize
       \caption{Panel (B): List of predictors for firms' failures.}
    \begin{footnotesize}
    \begin{tabularx}{\textwidth}{>{\raggedright\arraybackslash}p{8cm}>{\raggedright\arraybackslash}p{8cm}}
    \hline 
    \textbf{Variables} & \textbf{Description}\\  \midrule
    Interest Benchmarking (ZI) & It is a \textit{zombie} proxy proposed by \citet{caballero2008zombie} and calculated as $R^*=rs_{t-1}BS_{i,t-1}+(\frac{1}{5}\sum_{j=1}^{5}rl_{t-j})BL_{i,t-1} + rcb_{5y,t} \cdot Bonds_{i,t-1}$, where $BS_{i,t-1}$ are short-term bank loans, $BL_{i,t-1}$ are long-term bank loans, $rs_{t-1}$ are the average short-term prime rate in year $t$, $rl_{t-j}$ is the average long-term prime rate in year $t$, $Bonds$ are the total outstanding bonds, $rcb_{5y,t}$ is the minimum observed rate on any convertible corporate bond issued over the previous five years.\\
    
    Interest Coverage Ratio (ZI) &  It is calculated as EBIT/Interest Expenses. When it is less than one for three consecutive years and the firm is at least ten years old, then \citet{bankofkorea2013}, \citet{mcgowan2018walking} and \citet{banerjee2018rise} assume a firm is a \textit{zombie}.\\
    
    Financial Misallocation (ZI) &  It is a binary indicator adopted by \cite{schivardi2021credit} for catching \textit{zombie lending}, based on both $ROA\frac{\frac{1}{3}\sum_{t=1}^{3}EBITDA_t}{Total\:Assets} < prime$ and $Leverage = \frac{Financial\:Debt}{Total\:Assets} > \tilde{L}$, where $prime$ is the measure of the cost of capital for firms with a Z-score equal to 1 or 2, and where $\tilde{L}$ is the median value of leverage in the current year for firms that exited in two following years.\\
    
    Negative value added (ZI) & It is a binary variable to spot\textit{zombie firms} \citep{mcgowan2018walking}, equal to one when the value added is negative, i.e. when the value of sold output is less than purchases of intermediate inputs.\\
    
    Profitability (ZI) & Calculated as $EBITDA/Total\:Assets$, and adopted by \citet{schivardi2021credit} as a control for \textit{zombie lending} \\
    
    Financial Constraints (FC) & It is a proxy of financial constraints as in \cite{nickell1999does}, calculated as a ratio between interest payments and cash flow\\
    
    Size-Age (FC) & It is a synthetic indicator proposed by \cite{hadlock2010new}, equal to $-0.737*\log (total assets) + 0.043*\log(total assets))^2 - 0.040 * age$.\\
    
    Financial Sustainability (FC) & It is a ratio calculated as Financial Expenses over Operating Revenues.\\
    
    Capital Adequacy Ratio (FC) &  It is a ratio of Shareholders' Funds over Short and Long Term Debts.\\
    
    Liquidity Ratio (FA) & (Current/Assets - Stocks)/Current/Liabilities\\
    
    Solvency Ratio (FA) & (Shareholders funds / (Non current liabilities + Current liabilities)) * 100 \\
    
    Liquidity Returns (FA) & It is the ratio of cash flow over total asset\\
    
    Tax and Pension Payables (FA) & It is the ratio of the sum of tax and pension payables over total assets.\\ \hline
\end{tabularx}
\end{footnotesize}
\end{minipage}
\end{table}

\begin{table}[H]
\centering
\caption{Missing predictors and firms' failures - Chi-square tests\label{tab:chi-squared}} 
\resizebox{0.8\textwidth}{!}{%
\begin{tabular}{lccc}
\toprule
\addlinespace
& \multicolumn{2}{c}{\textit{Firm's failure}}               \\ \cline{2-3} 
\multicolumn{1}{l}{}&\multicolumn{1}{c}{0}&\multicolumn{1}{c}{1}&\multicolumn{1}{c}{Test Statistic}\tabularnewline
&\multicolumn{1}{c}{{\scriptsize $N=287587$}}&\multicolumn{1}{c}{{\scriptsize $N=17319$}}&\tabularnewline
\midrule
Interest Benchmarking~:~0&38\%~{\scriptsize~~(110524)}&61\%~{\scriptsize~~(10530)}&$\chi^{2}_{1}$=3414.25,~P\textless 0.001\tabularnewline \vspace{0.15cm}
Interest Benchmarking~:~1&62\%~{\scriptsize~~(177063)}&39\%~{\scriptsize~~(6789)}&\tabularnewline
Interest Coverage Ratio~:~0&37\%~{\scriptsize~~(105907)}&49\%~{\scriptsize~~(8422)}&$\chi^{2}_{1}$=970.93,~P\textless 0.001\tabularnewline \vspace{0.15cm}
Interest Coverage Ratio~:~1&63\%~{\scriptsize~~(181680)}&51\%~{\scriptsize~~(8897)}&\tabularnewline
Negative value added~:~0&34\%~{\scriptsize~~(~98014)}&63\%~{\scriptsize~~(10915)}&$\chi^{2}_{1}$=5958.81,~P\textless 0.001\tabularnewline \vspace{0.15cm}
Negative value added~:~1&66\%~{\scriptsize~~(189573)}&37\%~{\scriptsize~~(6404)}&\tabularnewline
Financial Constraint~:~0&37\%~{\scriptsize~~(105904)}&49\%~{\scriptsize~~(8419)}&$\chi^{2}_{1}$=968.27,~P\textless 0.001\tabularnewline \vspace{0.15cm}
Financial Constraint~:~1&63\%~{\scriptsize~~(181683)}&51\%~{\scriptsize~~(8900)}&\tabularnewline
Financial Misallocation~:~0&39\%~{\scriptsize~~(112560)}&54\%~{\scriptsize~~(9276)}&$\chi^{2}_{1}$=1415.82,~P\textless 0.001\tabularnewline \vspace{0.15cm}
Financial Misallocation~:~1&61\%~{\scriptsize~~(175027)}&46\%~{\scriptsize~~(8043)}&\tabularnewline
Total Factor Productivity~:~0&36\%~{\scriptsize~~(104345)}&38\%~{\scriptsize~~(6600)}&$\chi^{2}_{1}$=23.52,~P\textless 0.001\tabularnewline \vspace{0.15cm}
Total Factor Productivity~:~1&64\%~{\scriptsize~~(183242)}&62\%~{\scriptsize~~(10719)}&\tabularnewline
Solvency Ratio~:~0&41\%~{\scriptsize~~(118851)}&63\%~{\scriptsize~~(10897)}&$\chi^{2}_{1}$=3115.5,~P\textless 0.001\tabularnewline \vspace{0.15cm}
Solvency Ratio~:~1&59\%~{\scriptsize~~(168736)}&37\%~{\scriptsize~~(6422)}&\tabularnewline 
Liquidity Ratio~:~0&42\%~{\scriptsize~~(119357)}&72\%~{\scriptsize~~(12543)}&$\chi^{2}_{1}$=6362.72,~P\textless 0.001\tabularnewline \vspace{0.15cm}
Liquidity Ratio~:~1&58\%~{\scriptsize~~(168230)}&28\%~{\scriptsize~~(4776)}&\tabularnewline
Size-Age~:~0&42\%~{\scriptsize~~(120260)}&75\%~{\scriptsize~~(12989)}&$\chi^{2}_{1}$=7310.19,~P\textless 0.001\tabularnewline \vspace{0.15cm}
Size-Age~:~1&58\%~{\scriptsize~~(167327)}&25\%~{\scriptsize~~(4330)}&\tabularnewline
Liquidity Returns~:~0&39\%~{\scriptsize~~(112561)}&54\%~{\scriptsize~~(9277)}&$\chi^{2}_{1}$=1416.88,~P\textless 0.001\tabularnewline \vspace{0.15cm}
Liquidity Returns~:~1&61\%~{\scriptsize~~(175026)}&46\%~{\scriptsize~~(8042)}&\tabularnewline
Labour Productivity~:~0&34\%~{\scriptsize~~(~97253)}&36\%~{\scriptsize~~(6221)}&$\chi^{2}_{1}$=32.23,~P\textless 0.001\tabularnewline \vspace{0.15cm}
Labour Productivity~:~1&66\%~{\scriptsize~~(190334)}&64\%~{\scriptsize~~(11098)}&\tabularnewline
Profitability~:~0&37\%~{\scriptsize~~(105907)}&49\%~{\scriptsize~~(8422)}&$\chi^{2}_{1}$=970.93,~P\textless 0.001\tabularnewline \vspace{0.15cm}
Profitability~:~1&63\%~{\scriptsize~~(181680)}&51\%~{\scriptsize~~(8897)}&\tabularnewline
Financial Sustainability~:~0&41\%~{\scriptsize~~(117294)}&64\%~{\scriptsize~~(11119)}&$\chi^{2}_{1}$=3673.94,~P\textless 0.001\tabularnewline \vspace{0.15cm}
Financial Sustainability~:~1&59\%~{\scriptsize~~(170293)}&36\%~{\scriptsize~~(6200)}&\tabularnewline
Capital Intensity~:~0&36\%~{\scriptsize~~(104122)}&42\%~{\scriptsize~~(7325)}&$\chi^{2}_{1}$=261.17,~P\textless 0.001\tabularnewline \vspace{0.15cm}
Capital Intensity~:~1&64\%~{\scriptsize~~(183465)}&58\%~{\scriptsize~~(9994)}&\tabularnewline
\bottomrule
\end{tabular}}
		\begin{tablenotes}
		\footnotesize
		\singlespacing
        \item Note: Chi-square tests for the null hypothesis that missing predictors do not correlate with the event of failure. Number of observations in parentheses.  
 \end{tablenotes}
\end{table}

\begin{figure}[H] 
  \caption{\review{Patterns of missingness over time across firms}} 
  \label{fig:non-failing}
\begin{subfigure}{\textwidth}
\centering
   \includegraphics[width=0.6\linewidth]{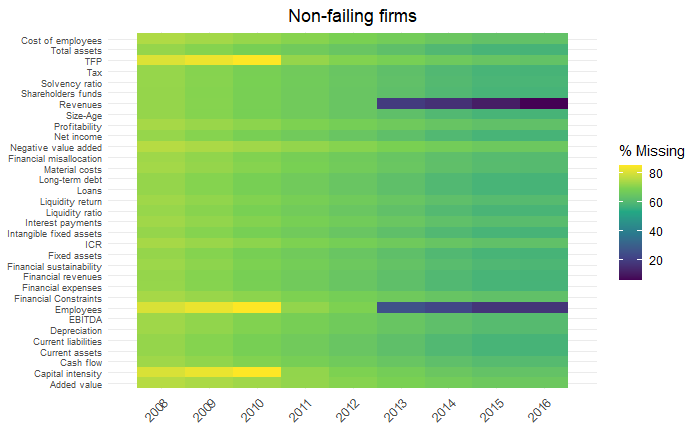}
   \caption*{\review{Panel (A): Share of missing values out of 287,787 non-failing firms.}}
\end{subfigure}
\begin{subfigure}{\textwidth}
\centering
   \includegraphics[width=0.6\linewidth]{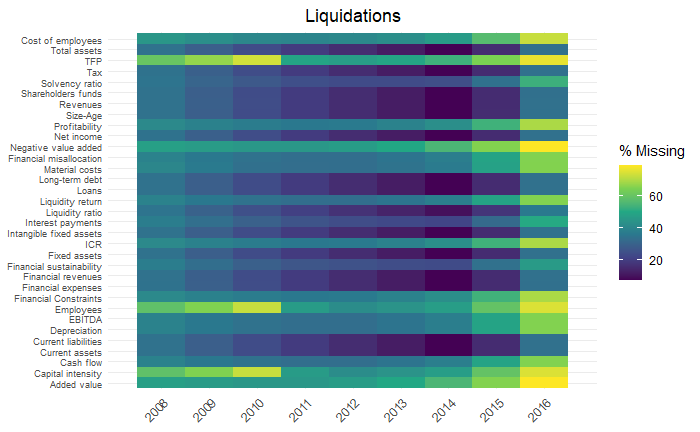}
   \caption*{\review{Panel (B): Share of missing values out of 7,221 firms in liquidation.}}
\end{subfigure}
\begin{subfigure}{\textwidth}
\centering
   \includegraphics[width=0.6\linewidth]{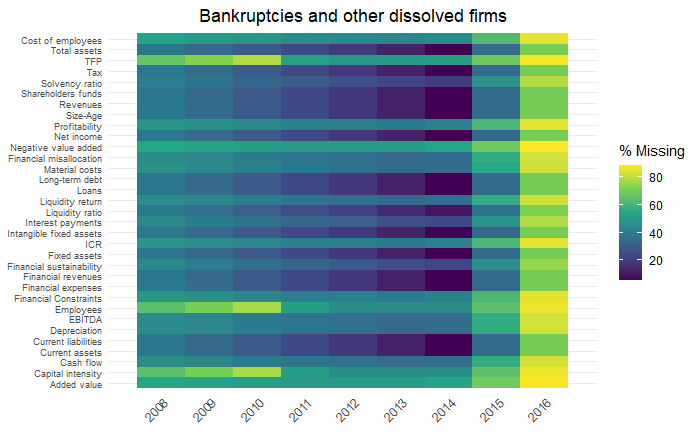}
   \caption*{\review{Panel (C): Share of missing values out of 4,718 bankruptcies and other dissolved firms.}}
\end{subfigure}
\end{figure}

\begin{table}[H]
\centering
\caption{\review{\textit{Zombies}, geography and selected financial indicators}}
\label{tab:zombies_geography_table}
\captionsetup{font=footnotesize}
 \resizebox{\textwidth}{!}{%
     \begin{tabular}{llcccccc}
    \hline
    & & \multicolumn{3}{c}{\textbf{Interest payments}} & \multicolumn{3}{c}{\textbf{Value generation}} \\
    \cline{3-8}
    Region & NUTS 2 & Common support & Zombies & ICR$<1$ & Common support & Zombies & Value added$<0$ \\ \hline
    Abruzzo & ITF1  & 0.16 & 0.48 & 0.36 & 0.15 & 0.57 & 0.27 \\ 
    Basilicata & ITF5  & 0.19 & 0.53 & 0.29 & 0.12 & 0.66 & 0.22 \\ 
    Calabria &   ITF6  & 0.20 & 0.50 & 0.30 & 0.13 & 0.64 & 0.23 \\ 
    Campania &   ITF3  & 0.16 & 0.57 & 0.27 & 0.12 & 0.62 & 0.25 \\ 
    Emilia-Romagna & ITH5  & 0.14 & 0.29 & 0.57 & 0.14 & 0.46 & 0.39 \\ 
    Friuli-Venezia Giulia  & ITH4  & 0.15 & 0.26 & 0.59 & 0.14 & 0.49 & 0.37 \\ 
    Lazio & ITI4  & 0.15 & 0.55 & 0.31 & 0.13 & 0.62 & 0.25 \\ 
    Liguria &  ITC3  & 0.20 & 0.37 & 0.43 & 0.14 & 0.54 & 0.32 \\ 
    Lombardia &  ITC4  & 0.15 & 0.28 & 0.57 & 0.14 & 0.48 & 0.38 \\ 
    Marche &  ITI3  & 0.17 & 0.41 & 0.42 & 0.16 & 0.53 & 0.31 \\ 
    Molise &  ITF2  & 0.13 & 0.46 & 0.41 & 0.15 & 0.57 & 0.29 \\ 
    Piemonte & ITC1  & 0.15 & 0.29 & 0.56 & 0.15 & 0.46 & 0.39 \\ 
    Puglia & ITF4  & 0.19 & 0.49 & 0.33 & 0.15 & 0.58 & 0.26 \\ 
    Sardegna &  ITG2  & 0.17 & 0.30 & 0.52 & 0.15 & 0.54 & 0.31 \\ 
    Sicilia & ITG1  & 0.18 & 0.41 & 0.41 & 0.14 & 0.53 & 0.33 \\ 
    Trentino-Alto Adige & ITH1/2  & 0.18 & 0.46 & 0.36 & 0.17 & 0.53 & 0.30 \\ 
    Toscana &  ITI1  & 0.16 & 0.25 & 0.59 & 0.12 & 0.42 & 0.46 \\ 
    Umbria &  ITI2  & 0.21 & 0.36 & 0.43 & 0.17 & 0.46 & 0.37 \\ 
    Valle d'Aosta &  ITC2  & 0.06 & 0.28 & 0.66 & 0.14 & 0.45 & 0.41 \\ 
    Veneto &  ITH3  & 0.15 & 0.30 & 0.54 & 0.15 & 0.45 & 0.40 \\ \hline
Num. obs. & &  \multicolumn{3}{c}{30,380} & \multicolumn{3}{c}{24,351} \\ \hline
    \end{tabular}}
\singlespacing
\caption*{\review{We report the proportions of \textit{zombies} against firms with ICR lower than one and against firms with negative value-added for each NUTS 2-digit region in Italy. Common support indicates the overlap between \textit{zombie} firms and the firms that have problems with interest payments ($ICR < 1$) and have a negative value-added, respectively. 
}}
\end{table}

\newpage
\begin{figure}[H]
\centering
  \caption{\review{\textit{Zombie firms} and industries}}
  \captionsetup{font=footnotesize}
  \label{fig:zombie_industries}
 \begin{minipage}{.5\textwidth}
  \centering
  \label{fig:ICR2}
  \includegraphics[width=0.9\linewidth]{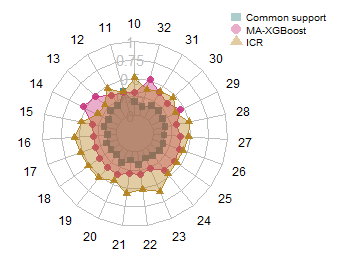}\par
  \subcaption{\textit{Zombies} and interest payments}
\end{minipage}%
\begin{minipage}{.5\textwidth}
  \centering
  \label{fig:NEG_VA2} \includegraphics[width=0.9\linewidth]{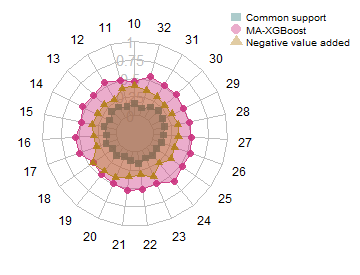}\par
   \subcaption{\textit{Zombies} and value destruction}
\end{minipage}
\footnotesize
\singlespacing
\caption*{\review{The rays of the radar show, at the industry level (NACE Rev. 2, 2-digit manufacturing codes), \textit{zombie} firms against firms that report Interest Coverage Ratio (ICR) of less than one (panel a) and the proportion of firms reporting negative value added (panel b). The square nodes indicate the common areas where the segments overlap. The circles represent the fraction of \textit{zombies} that we detect with MA-XGBoost. The triangles indicate the fraction of firms identified with $ICR 1$ or negative value added, alternatively. Along each ray, the values of the squares, circles, and triangles sum to one. Panel (a) and (b) include a total of 30,380 and 24,351 observations, respectively.}}
\end{figure}

\newpage
\begin{table}[H]
\centering
\caption{\review{\textit{Zombies}, industries and viability indicators}}
\captionsetup{font=footnotesize}
\label{tab:zombies_industries_table}
 \resizebox{\textwidth}{!}{%
     \begin{tabular}{llcccccc}
    \hline
  &  & \multicolumn{3}{c}{\textbf{Interest payments}} & \multicolumn{3}{c}{\textbf{Value generation}} \\
    \cline{3-8}
        Manufacturing of & NACE & Common support & Zombies & ICR$<1$ & Common support & Zombies & Value added$<0$ \\ \hline
           Food products  & 10   & 0.18 & 0.31 & 0.50 & 0.15 & 0.45 & 0.40 \\
           Beverages  & 11   & 0.33 & 0.33 & 0.33 & 0.12 & 0.50 & 0.38 \\ 
           Tobacco products  & 12   & 0.19 & 0.35 & 0.46 & 0.16 & 0.55 & 0.29\\ 
           Textiles  & 13   & 0.19 & 0.49 & 0.31 & 0.17 & 0.51 & 0.32 \\ 
           Wearing apparel  & 14   & 0.15 & 0.53 & 0.32 & 0.13 & 0.55 & 0.32 \\ 
           Leather  & 15   & 0.17 & 0.34 & 0.49 & 0.17 & 0.49 & 0.34 \\ 
           Wood and cork products  & 16   & 0.12 & 0.31 & 0.57 & 0.13 & 0.56 & 0.31 \\ 
           Paper products & 17   & 0.15 & 0.32 & 0.53 & 0.16 & 0.54 & 0.30 \\ 
           Printing and reproduction of recorded media  & 18   & 0.14 & 0.34 & 0.52 & 0.11 & 0.44 & 0.45 \\ 
           Coke and refined petroleum products  & 19   & 0.13 & 0.34 & 0.53 & 0.14 & 0.46 & 0.41\\ 
           Chemicals and chemical products  & 20   & 0.19 & 0.36 & 0.45 & 0.10 & 0.49 & 0.41 \\ 
           Pharmaceutical products and preparations  & 21   & 0.12 & 0.30 & 0.57 & 0.13 & 0.53 & 0.34 \\ 
           Rubber and plastic products  & 22   & 0.18 & 0.31 & 0.52 & 0.17 & 0.51 & 0.32\\ 
           Other non  metallic mineral products  & 23   & 0.12 & 0.26 & 0.61 & 0.12 & 0.49 & 0.39 \\ 
           Basic metals  & 24   & 0.14 & 0.39 & 0.46 & 0.14 & 0.59 & 0.27 \\ 
           Fabricated metal products  & 25   & 0.16 & 0.41 & 0.44 & 0.11 & 0.53 & 0.35 \\ 
           Computer, electronic and optical products  & 26   & 0.16 & 0.39 & 0.44 & 0.14 & 0.54 & 0.32 \\ 
           Electrical equipment  & 27   & 0.15 & 0.37 & 0.48 & 0.14 & 0.52 & 0.34 \\ 
           Machinery and equipment n.e.c.  & 28   & 0.17 & 0.33 & 0.50 & 0.16 & 0.50 & 0.35\\ 
           Motor vehicle, trailers and semi-trailers  & 29   & 0.18 & 0.45 & 0.37 & 0.18 & 0.49 & 0.32 \\ 
           Other transport equipment  & 30   & 0.19 & 0.35 & 0.46 & 0.19 & 0.52 & 0.30 \\ 
           Forniture  & 31   & 0.19 & 0.40 & 0.41 & 0.16 & 0.51 & 0.33 \\ 
           Other manufacturing  & 32   & 0.12 & 0.51 & 0.37 & 0.10 & 0.55 & 0.35 \\ \hline
Num. obs. & &  \multicolumn{3}{c}{30,380} & \multicolumn{3}{c}{24,351} \\ \hline
    \end{tabular}}
\singlespacing
\caption*{\review{We report for 2-digit industries (NACE Rev. 2) the shares of \textit{zombies} compared to firms with an ICR of less than one and compared to firms with negative value added. The common column contains the proportions of firms classified as zombies by our algorithm that have either an ICR of less than one or negative value added.
}}
\end{table}

\newpage
\review{
\section{Methodologies} \label{appendix:methods}
\label{sec}
\subsection{XGBoost}
XGBoost is a gradient- boosting algorithm introduced by \cite{chen2016xgboost}. The algorithm uses a standard boosting method where $J$ decision trees are sequentially created to approximate the outcome. Each tree uses the information learned from the previous trees, and the final model can be expressed as follows:
$$
Y_{i,t}  = \sum_{j=1}^J \mathcal{T}_j\left(\bX_{i,t-1}; \mathcal{D}_j, \mathcal{W}_j \right) + \epsilon_{i,t-1}
$$
where $\mathcal{T}_j\left(\bX_{i,t-1}; \mathcal{D}_j, \mathcal{W}_j\right)$ corresponds to an independent tree with structure $\mathcal{D}_j$ and leaf weights $\mathcal{W}_j$. Note that $\epsilon_{i,t-1}$ is typically assumed to be zero-mean, but no probabilistic assumptions are made about it. The model approximation is built additively, minimizing the loss function iteratively. The loss function includes a regularization term to penalize the complexity of the model and avoid overfitting, and has the following form:
\begin{eqnarray*}
\mathcal{L} & = & \sum_{i=1}^N L\left(\hat{Y}_{i,t}, Y_{i,t}\right)+\sum_{j=1}^J \Omega\left(\mathcal{T}_j\right) \\ 
\Omega(\mathcal{T}_j) & = & \gamma T_j+\frac{1}{2} \lambda\|\mathcal{W}_j\|^2
\end{eqnarray*}
where $T_j$ and $\mathcal{W}_j$ represent the number and weights of the leaves of the $j$-th tree, respectively, while $\gamma$ and $\lambda$ are regularization parameters used to reduce complexity and avoid overfitting. 

XGBoost is characterized by the fact that it includes several tools that increase the training speed and deal with sparsity in the data \citep{bentejac2021comparative}. In traditional algorithms for splitting data, finding the best splits is quite time consuming. To be efficient, each variable is sorted and all potential splits are explored to determine the optimal one. XGBoost overcomes this problem by using a compressed column-based structure with parallel processing to avoid repeated sorting of the data. The algorithm for finding the best split is now based on the percentiles of the features, so only a subset of the possible splits is examined, which significantly increases the training speed. In addition, XGBoost uses \textit{default directions} to handle sparsity patterns. For missing values, if a feature required for a split is missing, XGBoost assigns the observation to a default direction (left or right) learned from the data. This approach is useful for handling and incorporating information about missing values into the model.

\newpage
\subsection{BART-MIA}
BART is a Bayesian sum-of-trees ensemble algorithm and its approach is based on a fully Bayesian probability model \citep{kapelner2015prediction, JSSv070i04}. The revised version of the BART model we use for our predictions can be expressed as follows:
\begin{equation}\label{tree}  \centering 
    Y_{i,t} = \sum_{j=1}^{J} \mathcal{T}_j(\bX_{i,t-1}; \mathcal{D}_j, \mathcal{M}_j) + \epsilon_{i,t-1}, \:\:\:\:\:\:\:\:\: \epsilon_{i, t-1} \sim \mathcal{N} (0, \sigma^2),
\end{equation}
where each $\mathcal{T}_j(\bX_{i,t-1}; \mathcal{D}_j, \mathcal{M}_j)$ denotes a unique binary tree. Each $\mathcal{T}_j$ is a function that sorts each unit into one of the sets of $m_j$ terminal nodes associated with mean parameters $\mathcal{M}_j \{\mu_1, ..., \mu_{m_j}\}$ based on a set of decision rules, $\mathcal{D}_j$. The error terms $\epsilon_{i, t-1}$ are usually assumed to be independent and identically normally distributed if the outcome is continuous \citep{chipman2010bart}.  
The Bayesian component of the algorithm is incorporated in a set of three different priors on: (i) the structure of the trees, $\mathcal{D}_j$ (this prior aims to limit the complexity of each tree $\mathcal{T}$ and serves as a regularization tool); (ii) the distribution of the outcome in the nodes, $\mathcal{M}_j$ (this prior aims to reduce the node predictions to the center of the distribution of the response variable $Y$); (iii) the error variance $\sigma^2$ (which bounds away $\sigma^2$ from very small values that would cause the algorithm to overfit the training data)\footnote{The choice of priors and the derivation of posterior distributions are discussed in detail in \cite{chipman2010bart} and \cite{JSSv070i04}. Namely: (i) the prior on the probability that a node splits at depth $k$ is $\beta(1k)^{-\eta}$, where $\beta \in (0,1), \eta \in [0, \infty)$ (these hyperparameters are generally chosen to be $\eta2$ and $\beta 0.95$); (ii) the prior on the probability distribution in the nodes is a normal distribution with zero mean: $\mathcal{N}(0, \sigma^2_q)$ where $\sigma_q \sigma_0/\sqrt{q}$ and $\sigma_0$ can be used to calibrate the plausible range of the regression function; (iii) the prior on the error variance is $\sigma^2 \sim InvGamma(v/2, v\lambda/2)$ where $\lambda$ is determined from the data such that the BART improves the RMSE of an OLS model in 90\% of cases.}. The goal of these priors is to regularize the algorithm and prevent individual trees from dominating the overall fit of the model. This property is considered important to balance the tendency of tree-based methods to overfit the training data \citep{JSSv070i04}.

BART-MIA extends the original BART algorithm by including additional information from patterns of missing values \citep{kapelner2015prediction}. This is done by introducing the possibility to share on a feature for missing values in each binary tree component of the BART algorithm, $\mathcal{T}$. Figure \ref{fig:mia} illustrates the intuition behind the MIA procedure in a simple case with only one variable (\textit{male}). In the presence of missing values in this variable, the algorithm creates a new feature (\textit{missing}) -- namely, an indicator variable that takes the value 1 when the $i$th observation for the variable is missing -- and uses it to perform the data split. As shown in \ref{fig:mia}, the algorithm in this case has three possible splits. As in the usual CART procedure, the algorithm chooses the one that minimizes the prediction error in the generated leaves $l_1$ and $l_2$.

As shown by \citet{twala2008good}, this splitting rule allows trees to better capture the direct influence of missing values as another predictor of the response variable. Furthermore, after a theoretical and empirical comparison of different missing value strategies in trees, \cite{josse2019consistency} show that MIA can handle both non-informative and informative missing values.
        \begin{figure}[H]
        \caption{\review{MIA splitting rules}}
		\begin{tikzpicture}[level distance=80pt, sibling distance=30pt, edge from parent path={(\tikzparentnode) -- (\tikzchildnode)}]
		\tikzset{every tree node/.style={align=center}}
		\Tree [.\node[rectangle,draw]{Total sample}; \edge node[auto=right,pos=.6]{Missing$=1$}; \node[circle,draw]{$\mu_{l_1}$}; \edge node[auto=left,pos=.6]{Control$=1$}; \node[circle,draw]{$\mu_{l_2}$};]
		\end{tikzpicture}
		\begin{tikzpicture}[level distance=80pt, sibling distance=30pt, edge from parent path={(\tikzparentnode) -- (\tikzchildnode)}]
		\tikzset{every tree node/.style={align=center}}
		\Tree [.\node[rectangle,draw]{Total sample}; \edge node[auto=right,pos=.6]{Control$=0$}; \node[circle,draw]{$\mu_{l_1}$}; \edge node[auto=left,pos=.6]{Missing$=1$}; \node[circle,draw]{$\mu_{l_2}$};]
		\end{tikzpicture}
		\begin{tikzpicture}[level distance=80pt, sibling distance=30pt, edge from parent path={(\tikzparentnode) -- (\tikzchildnode)}]
		\tikzset{every tree node/.style={align=center}}
		\Tree [.\node[rectangle,draw]{Total sample}; \edge node[auto=right,pos=.6]{Missing$=0$}; \node[circle,draw]{$\mu_{l_1}$}; \edge node[auto=left,pos=.6]{Missing$=1$}; \node[circle,draw]{$\mu_{l_2}$};]
		\end{tikzpicture}
        \singlespacing
        \caption*{\review{The three potential trees from the MIA procedure in a simple case with only one binary variable ( Control $\in \{0,1\}$).}}
        \label{fig:mia}
	\end{figure}
}

\newpage

\section{Selection of predictors with LASSO}\label{appendix:lasso}

In our analysis, we asserted that we need as much in-sample information as possible to reduce out-of-sample prediction errors. However, especially in small sample analyses, overfitting problems can arise as the dimensionality of the inputs increases. The so-called \textit{curse of dimensionality} is an obstacle when working with finite data samples and many variables. The basic reference is the work of \cite{bellman_adaptive}, who introduced the notion of dimensionality reduction. \cite{bargagli2020simplicity} and \cite{bargagli2022simple} discuss the role of regularization and dimensionality reduction in the context of machine learning.In this Appendix, we show what happens when we sift through firm-level data and attempt to extract a set of predictors with the highest ability to detect financial distress. The natural candidate for reducing the dimensionality of a matrix of predictors is the LOGIT-LASSO \citep{ahrens2019lassopack}, whose functional form in a panel setting is the following:
\begin{eqnarray}\label{eq:LOGITLASSO}
\argmin_{{\beta} \in \mathbb{R}^{p}} \:\: \frac{1}{2N} \sum_{i=1}^N \left(y_{i,t}(x_{i,t-1}^T\beta) - log(1 + e^{(x_{i,t-1}^T\beta)})\right)^2 \:\:\:
{\rm subject\,\,to\,\,} \|{\beta}\|_1 \leq k.
\end{eqnarray}
where $y_{i,t}$ is a binary variable equal to one if a firm $i$ failed at time $t$ and zero otherwise. Each $x_{i,t-1}$ is a lagged predictor chosen in $\mathbb{R}^{p}$ at time $t-1$, while $\|{\beta}\|_1\sum_{j1}^p |\beta_j|$ and $k 0$. The constraint $\|{\beta}\|_1 \leq k$ limits the complexity of the model to avoid overfitting, and $k$ is chosen following \citet{ahrens2019lassopack} as the value that maximizes the Extended Bayesian Information Criteria \citep{chen2008extended}. We use the rigorous penalization introduced by \cite{belloni2016inference} to account for the possible presence of heteroskedastic, non-Gaussian, and cluster-dependent errors. Table \ref{tab:predictors_of_failure} shows the ten highest ranked predictors.

Although the number of predictors varies over time, up to a maximum of twenty-one characteristics in 2015, we find a core set of frequently selected predictors. This stable set includes indicators of financial distress (\textit{Liquidity Returns, Interest Coverage Ratio, Interest Benchmark, Financial Constraint}) and indicators of firms' core economic activities (\textit{Negative value added, Total Factor Productivity, Size-Age}). Apparently, firms controlled by parent companies (\textit{Corporate Control}) are less likely to fail in each period, i.e. the predictor enters the algorithm with a negative coefficient\footnote{For an in-depth analysis of the impact of corporate control on the performance of Italian firms, see \citet{riccaboni2021firm}.}. The same is true for the \textit{Dummy Trademarks} and the \textit{Dummy Patents} since it makes sense that intangible assets reduce the probability of exit.

Some of the top predictors we showed were used to measure either financial distress or \textit{zombie lending}. However, the rankings change over time, and we cannot discern a meaningful pattern in these changes. In this context, we cannot rely on a single indicator (or set of them) to derive predictions about defaults. If we did, we would have a higher rate of \textit{false positives} (when our forecasts suggest that a firm is at risk of failure, but it is not) and a higher rate of \textit{false negatives} (when forecasts incorrectly suggest that a firm in trouble deserves a loan).

Indeed, at this stage, we cannot rule out the possibility that a different ranking over time is due to better use of the newly acquired information from the in-sample information after new failures have been observed. It could also be that a change in rankings reflects a change in the business environment in which firms operate. We presume applications to different aggregates (countries, regions, industries) may result in different rankings. We conclude that it is better to keep the full battery of predictors, provided there is no dimensionality problem in the predictors when we can train with a large number of observed firm-level outcomes.

\begin{landscape}
\vspace*{27mm}

\begin{table}[H]
\centering
\doublespacing
\caption{Top 10 predictors for firms' failures - Results from a rigorous LOGIT-LASSO}
\label{tab:predictors_of_failure}
	\resizebox{1.3\textwidth}{!}{%
		\begin{tabular}{cccccccccc}
		\toprule
			Rank & 2017                         & 2016                    & 2015                             & 2014                             & 2013                             & 2012                             & 2011            & 2010  & 2009               \\
			\midrule
		    1                           & Liquidity Returns           & Negative value added & Negative value added &  Negative value added & Liquidity Returns  & Negative value-added    &   Negative value added & Negative value added  & Negative value added  \\
			2                           & Negative value added  & Liquidity Returns          & Corporate Control    &  Liquidity Returns         & Negative value added & Profitability  &  Liquidity Returns  & Liquidity Returns & Liquidity Returns \\
			3                           & Corporate Control      & Corporate Control     & Financial Constraint        & Solvency Ratio  & Solvency Ratio  & Financial Constraint &  Financial Constraint & Profitability & Financial Constraint       \\
			4                           & Interest Coverage Ratio          & Financial Constraint         & Interest Coverage Ratio        & Profitability &  Profitability & Corporate Control & Corporate Control    & Financial Constraint          & Profitability\\
			5                           & Financial Constraint         & Interest Coverage Ratio         & Profitability & Financial Constraint       & Corporate Control    & Solvency Ratio & Solvency Ratio     & Corporate Control      & Corporate Control \\
			6                           & Solvency Ratio     & Size-Age    &  Solvency Ratio   & Corporate Control   & Financial Constraint        & Interest Coverage Ratio & Size-age     & Solvency Ratio     & Solvency Ratio \\
			7                           & Size-age     & Solvency Ratio    & Size-age     & Size-age  & Size-Age   & Liquidity Returns & Interest Coverage Ratio          & Region (NUTS 2)         & Interest Coverage Ratio \\
			8                           & Profitability & Profitability &  Interest Benchmark        & Interest Coverage Ratio       & Interest Coverage Ratio       & Size-age & Financial Misallocation & Dummy Trademarks    & Dummy Trademarks\\
			9                           & Interest Benchmark          & Interest Benchmark          &  Liquidity Ratio  & TFP       & TFP        & Dummy Patents & Dummy Trademarks    & Interest Coverage Ratio          & Size-age\\
			10                          & Liquidity Ratio    & Liquidity Ratio    & Capital Intensity          & Liquidity Ratio & Dummy Patents & Dummy Trademarks & Dummy Patents   & Dummy Patents   & Capital Intensity \\
			\bottomrule
		\end{tabular}
		}
		\begin{tablenotes}
\footnotesize
\singlespacing
\item Rankings are determined after conducting a rigorous LOGIT-LASSO \citep{robustlasso, belloni2016post} each year on the entire battery of predictors described in Figure \ref{tab:table2}. Only the first ten selections are reported. The procedure selects a different number of predictors each year, up to a maximum of 21.
\end{tablenotes}
\end{table}
\end{landscape}


\newpage
\review{
\section{Imputation of missing predictors}\label{appendix:missing_imputation}
In this Appendix, we show the performance of machine learning algorithms when we impute the missing values of predictors using two possibilities known in the literature: Out-of-Range and Median imputation. In this way, we can use all the observations available in the dataset. We first conduct the analysis keeping all categories of firm failures and then exclude the cases of liquidations, since the latter can sometimes be due to reasons other than financial distress.
The training and test datasets include 238,148 and 59,537 observations in each iteration, respectively, when liquidations are excluded. On the other hand, when liquidations are included, the training and test data sets comprise 243,924 and 60,982 observations, respectively.
\subsection{Imputation of out-of-range values}
We impute all missing values in the dataset with an out-of-range value, i.e. $10^{20}$.  
\begin{table}[H]
\begin{subtable}[h]{\textwidth}
\caption{Liquidations exluded}
\centering
    \begin{tabular}{lcccccc}
\hline 
 Method &  ROC  &  PR  &  F1-Score  &  BACC  &  $R^2$  &  Time \\ \hline \hline 
\textit{Logit}  &  0.9657  &  0.6689  &  0.1270 &  0.7588  &  0.4558  &  56.49 \\
\textit{Ctree}  &  0.9712  &  0.6999  &  0.1323  &  0.7686  &  0.4840  &  1628.08 \\ 
\textit{Random Forest}  &  0.9343  &  0.7000  &   0.2011   &   0.8371  &   0.4999   &   608.33  \\ 
\textit{XGBoost} &  0.9785  &  0.7592  &  0.1270  &  0.7586  &  0.5427  &  40.28    \\
\textit{BART}   &   0.9770  &  0.7482  &  0.1296  & 0.7635   &  0.5327  & 2736.74   \\ 
\textit{Super Learner}  &  0.9788  &  0.7574  &  0.1289  &  0.7626  &  0.5454  &  11765.18   \\ 
\hline  
\multicolumn{1}{c}{}
\end{tabular}
\end{subtable}
\begin{subtable}[h]{\textwidth}
\caption{Liquidations included}
\centering
    \begin{tabular}{lcccccc}
\hline 
\multicolumn{1}{p{\widthof{FT-XGBoost}}}{Method} &  AUC  &  PR  &  F1-Score  &  BACC  &  $R^2$  &  Time \\ \hline \hline
\textit{Logit} &  0.9496  &  0.6482  &  0.2039  &  0.7647  &  0.4173  &  76.89  \\ 
\textit{Ctree} &  0.9652 & 0.7230 & 0.2264  & 0.7919  &  0.48907   &  2835.230  \\
\textit{Random Forest} &   0.9410   &   0.7313   &   0.2680  &   0.8214   &   0.5130   &   695.48   \\
\textit{XGBoost} & 0.9733   &    0.7815  &  0.2039  &  0.7648  &  0.5525  &  45.29    \\
\textit{BART} &  0.9719 & 0.7718 & 0.2048 & 0.7659 & 0.5412 & 4056.36 \\
\textit{Super Learner} & 0.9748 & 0.7896 & 0.2039 & 0.7649 & 0.5620 & 13992.57 \\
\hline 
\end{tabular}
\end{subtable}
\caption{\review{Performance of the model with out-of-range values imputed data. All algorithms were trained with five-fold cross-validation.
All metrics correspond to the five-fold average. Time indicates the average seconds taken to train the model in each fold.}}
\end{table}}

\newpage
\review{
\subsection{Median imputation}
We impute the missing values of a variable $q$ with its median $q_{.5}$.
\begin{table}[H]
\begin{subtable}[h]{\textwidth}
\caption{Liquidations exluded}
\centering
    \begin{tabular}{lcccccc}
\hline 
\multicolumn{1}{p{\widthof{Random Forest}}}{Method} &  ROC  &  PR  &  F1-Score  &  BACC  &  $R^2$  &  Time \\ \hline \hline 
\textit{Logit} &  0.8825  &  0.2690 &  0.1465  &  0.7540  &  0.1495  &  50.77  \\ 
\textit{Ctree}  &  0.9270  &  0.4017  &  0.1492  &  0.7962  &  0.2364  &  1975.38  \\ 
\textit{Random Forest}   &  0.7767  &  0.4337  &  0.1487  &  0.7060  &  0.2667  &  855.37   \\ 
\textit{XGBoost} &   0.9584 & 0.5246 & 0.1271 & 0.7579 & 0.3281 & 39.51 \\
\textit{BART}   &  0.9485  &  0.4936  &  0.1328  &  0.7688  &  0.3014  &  4345.19    \\ 
\textit{Super Learner} &  0.9595  &  0.5393  &  0.1269  &  0.7576  &  0.3389  &  13927.68    \\ 
\hline  
\multicolumn{1}{c}{}  
\end{tabular}
\end{subtable}
\begin{subtable}[h]{\textwidth}
\caption{Liquidations included}
\centering
    \begin{tabular}{lcccccc}
\hline 
\multicolumn{1}{p{\widthof{Random Forest}}}{Method} &  ROC  &  PR  &  F1-Score  &  BACC  &  $R^2$  &  Time \\ \hline \hline 
\textit{Logit} & 0.8907 & 0.3966 & 0.2282 & 0.7667 & 0.2302 & 60.44 \\ 
\textit{Ctree} & 0.9237 & 0.5059 & 0.2187 & 0.7808 & 0.3082 & 2429.78 \\
\textit{Random Forest} & 0.8281 & 0.5453 & 0.2227 & 0.7304 & 0.3420 & 1167.71 \\
\textit{XGBoost} &  0.9536 & 0.6175 & 0.2037 & 0.7636 & 0.3942 & 40.93 \\
\textit{BART} & 0.9440 & 0.5890 & 0.2063 & 0.7663 & 0.3696 & 4702.54 \\
\textit{Super Learner} &  0.9551 & 0.6295 & 0.2043 & 0.7643 & 0.4051 & 15218.50 \\
\hline
\end{tabular}
\end{subtable}
\caption{\review{Performance of models with column-wise median imputed data. All algorithms are trained with five-fold cross-validation. 
All metrics correspond to the five-fold average. The time indicates the average seconds taken to train the model in each fold.}}
\end{table}}

\newpage

\end{document}